\documentclass[iop,apjl]{emulateapj}
\usepackage{natbib}
\usepackage{graphicx}
\usepackage{latexsym}
\usepackage{amsmath,amsfonts}
\usepackage{longtable}
\usepackage{multirow,booktabs}
\usepackage{hyperref}
\usepackage{afterpage} 
\usepackage[normalem]{ulem}
\usepackage{mathtools,xparse}

\slugcomment{Accepted for publication in ApJ, October 31, 2017}

\hypersetup{colorlinks=false,pdfborder={0 0 0}}

\setcitestyle{notesep={; }}

\DeclarePairedDelimiter{\norm}{\lVert}{\rVert}

\DeclarePairedDelimiterX{\infdivx}[2]{(}{)}{%
  #1\;\delimsize\|\;#2%
}
\newcommand{\infdiv}{D\infdivx}

\def\sgra{Sgr~A$^{\ast}$}

\def\lsim{\mathrel{\raise.3ex\hbox{$<$\kern-.75em\lower1ex\hbox{$\sim$}}}}
\def\gsim{\mathrel{\raise.3ex\hbox{$>$\kern-.75em\lower1ex\hbox{$\sim$}}}}
\def\gtwid{\mathrel{\raise.3ex\hbox{$>$\kern-.75em\lower1ex\hbox{$\sim$}}}}
\def\proptwid{\mathrel{\raise.3ex\hbox{$\propto$\kern-.75em\lower1ex\hbox{$\sim$}}}}
\newcommand\lrover[1]{\overset{\text{\tiny$\leftrightarrow$}}{\mathbf{#1}}}

\begin{document}

\title{ Dynamical Imaging with Interferometry }
\shorttitle{Dynamical Interferometric Imaging}

\author{Michael D.~Johnson\altaffilmark{1}, 
Katherine L.\ Bouman\altaffilmark{1,2}, 
Lindy Blackburn\altaffilmark{1}, 
Andrew A.\ Chael\altaffilmark{1}, 
Julian Rosen, 
Hotaka Shiokawa\altaffilmark{1}, 
Freek Roelofs\altaffilmark{3},
Kazunori Akiyama\altaffilmark{4}, 
Vincent~L.\ Fish\altaffilmark{4}, 
and Sheperd S.\ Doeleman\altaffilmark{1}
}
\shortauthors{Johnson et al.}
\altaffiltext{1}{Harvard-Smithsonian Center for Astrophysics, 60 Garden Street, Cambridge, MA 02138, USA}
\altaffiltext{2}{Massachusetts Institute of Technology, Computer Science and Artificial Intelligence Laboratory, 32 Vassar Street, Cambridge, MA 02139, USA}
\altaffiltext{3}{Department of Astrophysics, Institute for Mathematics, Astrophysics and Particle Physics, Radboud University, P.O.\ Box 9010, 6500 GL Nijmegen, The Netherlands}
\altaffiltext{4}{Massachusetts Institute of Technology, Haystack Observatory, Route 40, Westford, MA 01886, USA}
\email{mjohnson@cfa.harvard.edu}

\begin{abstract}
By linking widely separated radio dishes, the technique of very long baseline interferometry (VLBI) can greatly enhance angular resolution in radio astronomy. However, at any given moment, a VLBI array only sparsely samples the information necessary to form an image.  Conventional imaging techniques partially overcome this limitation by making the assumption that the observed cosmic source structure does not evolve over the duration of an observation, which enables VLBI networks to accumulate information as the Earth rotates and changes the projected array geometry. 
 Although this assumption is appropriate for nearly all VLBI, it is almost certainly violated for submillimeter observations of the Galactic Center supermassive black hole, Sagittarius~A$^\ast$ (\sgra), which has a gravitational timescale of only ${\sim}20$~seconds and exhibits intra-hour variability. To address this challenge, we develop several techniques to reconstruct dynamical images (``movies'') from interferometric data. Our techniques are applicable to both single-epoch and multi-epoch variability studies, and they are suitable for exploring many different physical processes including flaring regions, stable images with small time-dependent perturbations, steady accretion dynamics, or kinematics of relativistic jets. Moreover, dynamical imaging can be used to estimate time-averaged images from time-variable data, eliminating many spurious image artifacts that arise when using standard imaging methods. We demonstrate the effectiveness of our techniques using synthetic observations of simulated black hole systems and 7mm Very Long Baseline Array observations of M87, and we show that dynamical imaging is feasible for Event Horizon Telescope observations of \sgra. 
\end{abstract}


\section{Introduction}
\label{sec::Introduction}
  
Very long baseline interferometry (VLBI) provides exceptional angular resolution but only sparsely samples the Fourier components of an image. A powerful technique to enhance interferometric imaging utilizes the Earth's rotation -- as the Earth rotates, each baseline connecting an antenna pair tracks through, and samples, a range of image Fourier components \citep[see, e.g.,][]{Ryle_1962,Kellermann_Moran_2001,TMS}. In its conventional implementation, Earth rotation synthesis imaging assumes that the source being imaged is static over the observing duration (typically ${\sim}$hours). This assumption is reasonable for nearly all astrophysical sources of interest, although a few sources have shown detectable structural changes within a single observation \citep[e.g.,][]{Reid_2014}, most commonly through rapid swings of polarization angle \citep[e.g.,][]{Gabuzda_2000}. 

One notable case for which the static source assumption is likely to fail is the Galactic Center supermassive black hole, Sagittarius~A$^\ast$ (\sgra). Because \sgra\ has a mass of approximately $M \approx 4\times 10^6\,M_{\odot}$ \citep{Ghez_2008,Gillessen_2009}, its gravitational timescale is only $GM/c^3 \approx 20~{\rm seconds}$ and its innermost stable circular prograde orbits have periods of only $4{-}30$ minutes, depending on the black hole spin \citep{Bardeen_1972}. In terms of observed variability, \sgra\ regularly flares with ${\sim}$hour timescales \citep[e.g.,][]{Marrone_2008,Yusef-Zadeh_2009,Brinkerink_2015}, and its polarization shows intense variations on similar timescales \citep[e.g.,][]{Marrone_2006,Eckart_2006,Zamaninasab_2010,Johnson_2015}.

\begin{figure}[t]
\centering
\includegraphics*[width=\columnwidth]{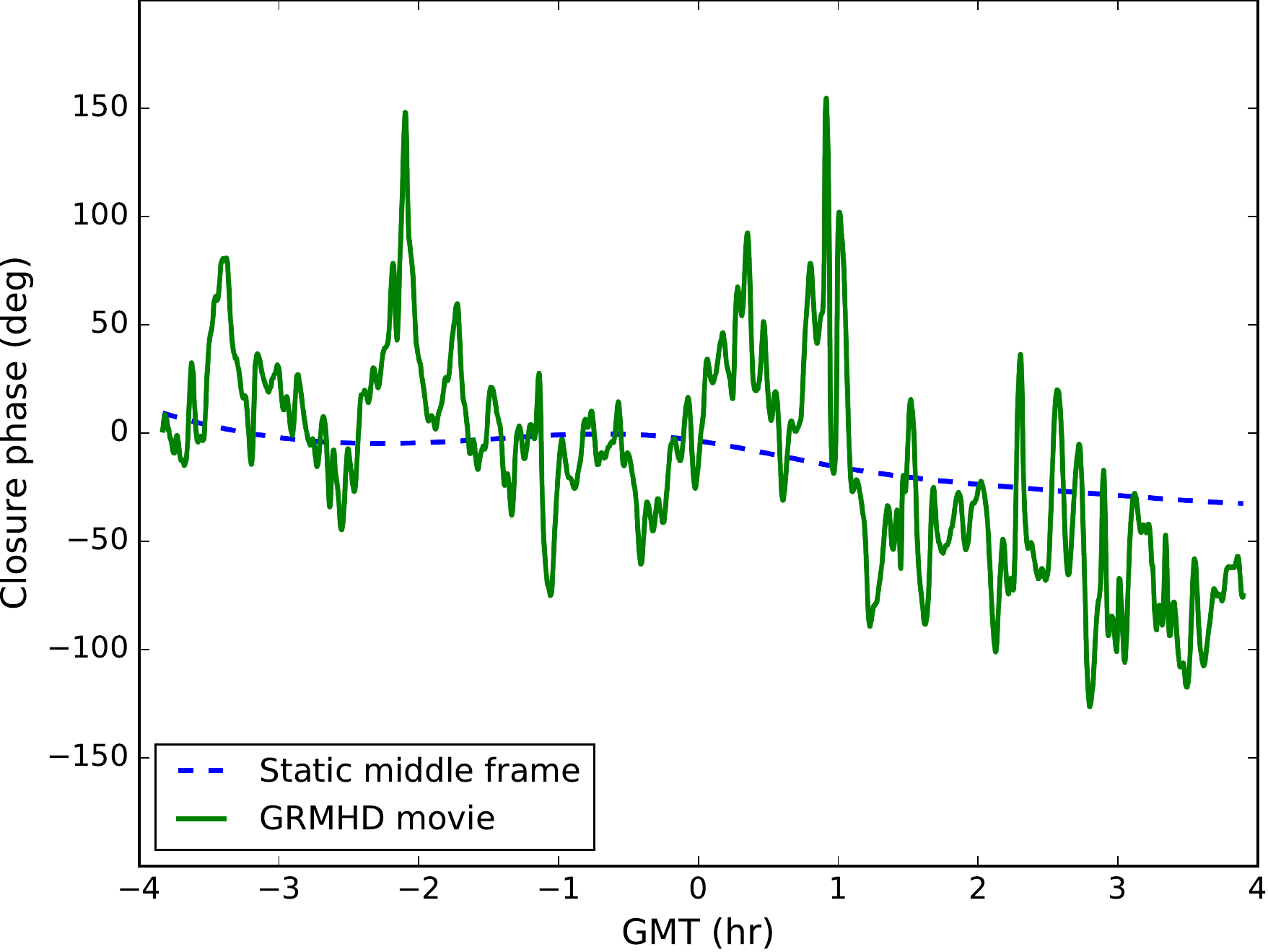}
\caption
{ 
VLBI phase measurement, the closure phase, over time for the SPT-LMT-ALMA triplet of EHT antennas using mock observations of a time variable general relativistic magnetohydrodynamic (GRMHD) simulation of \sgra\ \citep{Shiokawa_2013}.  The phases in blue show the array response to a single static frame in the GRMHD movie. The mild variations for this case reflect the Earth's rotation and show the modest additional information available to Earth-rotation synthesis.  The phases in green trace the array response to the full simulation, showing that the phase variations are dominated by intrinsic variability of the source. See \citet{Roelofs_2017} for additional examples and discussion.}
\label{fig::ClosurePhaseVariability}
\end{figure}

Until recently, limitations from optical depth and interstellar scattering have prevented studies of rapid structural variability of \sgra\ using VLBI \citep[e.g.,][]{Bower_2006,Lu_2011}. However, the advent of 1.3-mm VLBI with the Event Horizon Telescope (EHT) will soon permit imaging \sgra\ on spatial scales for which intrinsic variability may be significant \citep{Doeleman_2009}. Pronounced variability with accompanying structural change has already been seen in the polarization of \sgra\ with the EHT \citep{Johnson_2015}, although the total-intensity structure of \sgra\ has remained comparatively stable \citep{Doeleman_2008,Fish_2011,Fish_2016}. In addition to these observations and the short characteristic timescales of \sgra, numerical simulations suggest that conventional VLBI imaging techniques will be inapplicable for EHT observations of \sgra\ \citep[see Figure~\ref{fig::ClosurePhaseVariability} and, e.g.,][]{Broderick_Loeb_2006,Doeleman_2009,Dexter_2010,Baseline_Covariance,Lu_2016,Medeiros_2016b,Kim_2016,Medeiros_2016,Gold_2017,Roelofs_2017}. 

Nevertheless, interferometry provides capabilities to study rapidly varying structures. For example, using simulated observations of a ``hot spot'' orbiting \sgra\ \citep{Broderick_Loeb_2005,Broderick_Loeb_2006}, \citet{Doeleman_Hotspots} and \citet{Fish_Hotspots} demonstrated that robust VLBI observables can sensitively detect periodicities associated with these hot spots. More generally, \citet{Johnson_2014} showed that polarimetric VLBI enables microarcsecond astrometry of compact flaring structures, even for faint, non-periodic flares. Even conventional Earth rotation synthesis utilizes time-variable Fourier sampling to enhance imaging, and \citet{Baseline_Covariance} argued that intrinsic variability of a source can be exploited in the same way if the source variability can be modeled \citep[see, e.g.,][]{Sault_1997}. As Figure~\ref{fig::ClosurePhaseVariability} shows, while intrinsic variability of \sgra\ may readily break the static source assumption of conventional imaging, it also provides a rich source of information about the intrinsic variability.

In this paper, we develop techniques to reconstruct dynamical images (i.e., movies) from interferometric data. By accommodating intrinsic variability in the imaging procedure, we can study the dynamical activity of a source while avoiding spurious image features from the static-source assumption of conventional imaging algorithms. In a related approach, \citet{Lu_2016} have recently developed a prescription for scaling, averaging, and smoothing interferometric visibilities; the processed visibilities can then be imaged using standard VLBI imaging techniques.\footnote{This prescription is motivated by linearity of the Fourier transform: complex visibilities of the time-averaged image are equal to time-averaged visibilities of a variable image. See \citet{Shiokawa_2017} for generalized time-domain filtering of images.} They show that the resulting images are good approximations of the time-averaged image, especially when data from multiple observing epochs can be combined. Our focus is instead on reconstructing dynamical images of the time-variable source, while obtaining reliable approximations of the time-averaged image as a by-product. 

Our work is a generalization of the standard regularized minimization approach to VLBI imaging, which includes approaches such as the maximum entropy method \citep[MEM; see, e.g.,][]{Narayan_Nityananda_1986} and many other regularization functions. This approach, while not a strictly probabilistic model, can be motivated through a Bayesian framework wherein the minimization corresponds to maximizing the log posterior probability of a reconstructed image. In a separate paper, we explore an alternative approach to dynamical imaging via a modified Hidden Markov Model with a multivariate Gaussian image prior, and we derive closed-form expressions for both the maximum a posteriori image and its uncertainties \citep{Bouman_2017_StarWarps}.

We begin, in \S\ref{sec::Imaging_Background}, by reviewing the standard framework and procedure for VLBI imaging through regularized minimization, and we then generalize this framework to accommodate dynamical imaging. Next, in \S\ref{sec::Dynamical_Regularizers}, we develop three regularizers that can be used for dynamical imaging for a variety of scenarios. In \S\ref{sec::Interpolation}, we discuss using dynamical imaging for temporal interpolation. In \S\ref{sec::Imaging_Examples}, we show example results using simulated data and Very Long Baseline Array (VLBA) observations of M87, and in \S\ref{sec::Summary} we summarize our main results and conclusions.

\begin{figure}[t]
\centering
\includegraphics*[width=\columnwidth]{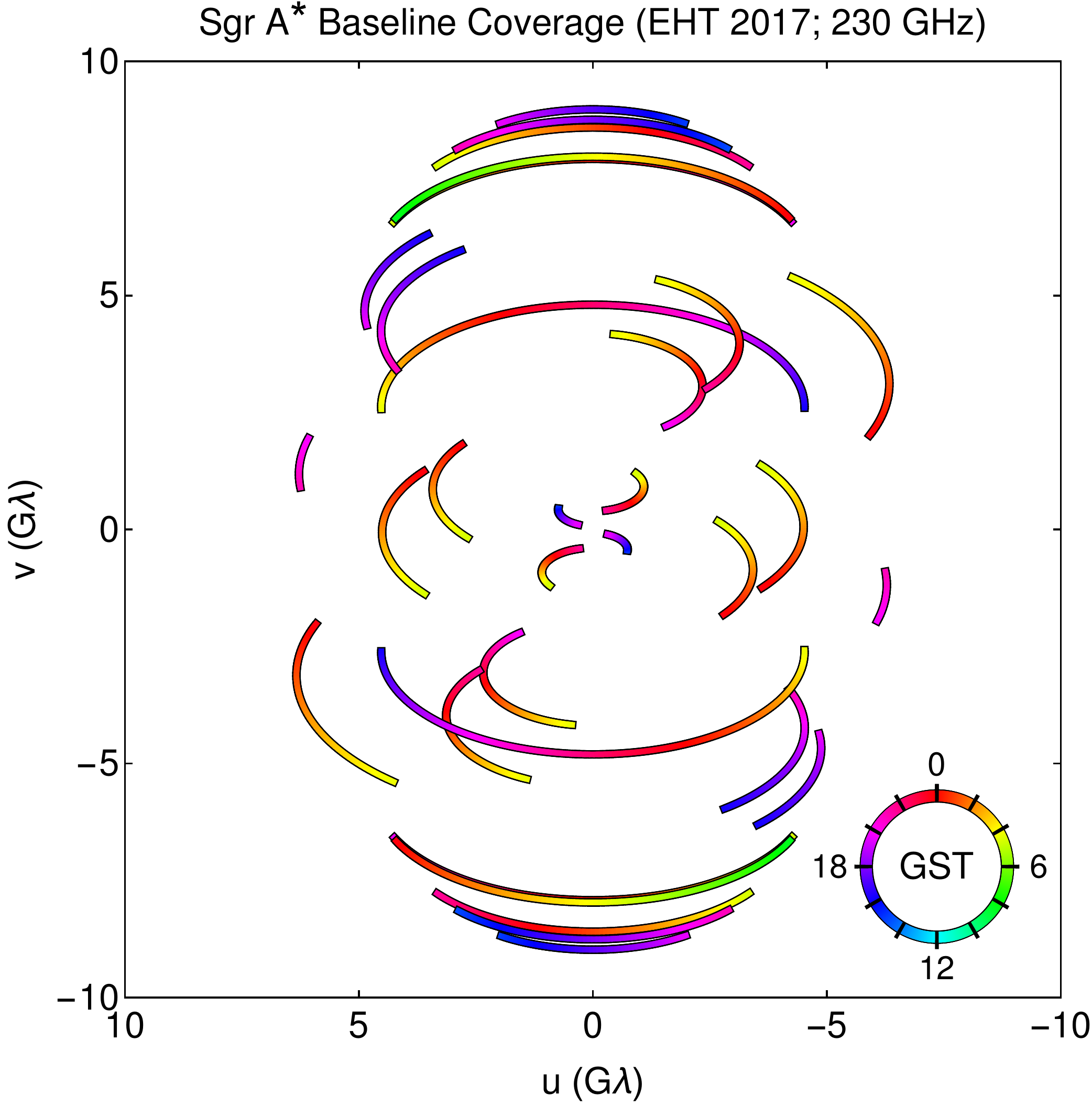}\\
\vspace{0.2cm}
\includegraphics*[width=\columnwidth]{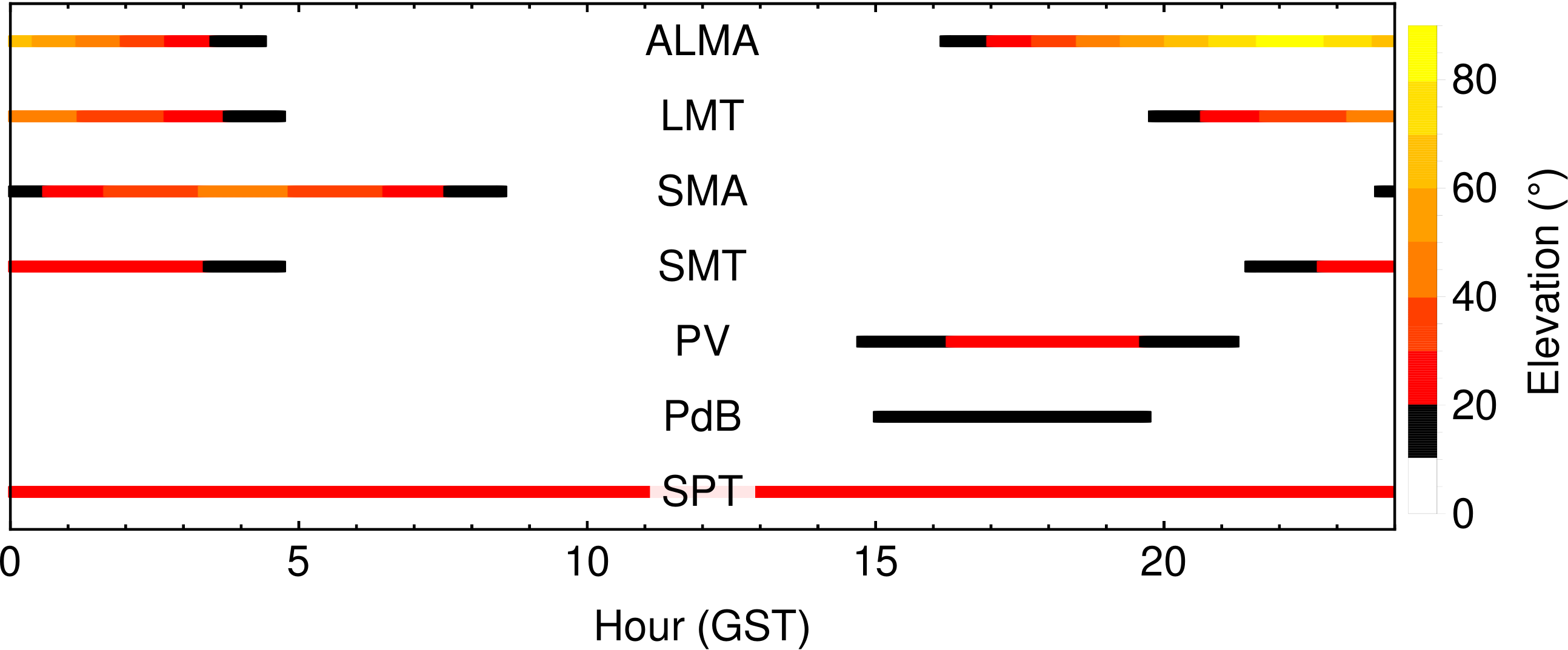}
\caption
{ 
(Top) Baseline coverage for the EHT observing \sgra. Baselines are colored by Greenwich Sidereal Time (GST) to indicate the snapshot u-v coverage at each time. (Bottom) Elevation of \sgra\ as a function of GST for each site, with a cutoff of $10^\circ$. Current EHT sites are the Atacama Large Millimeter/submillimeter Array (ALMA), the Large Millimeter Telescope (LMT), the Submillimeter Array (SMA), the Submillimeter Telescope (SMT), the Institut de Radioastronomie Millim\'{e}trique (IRAM) telescope on Pico Veleta (PV), the IRAM Plateau de Bure Interferometer (PdB), and the South Pole Telescope (SPT). Note that PdB did not participate in 2017 EHT observations.
}
\label{fig::u-v_Coverage}
\end{figure}


\section{Fundamentals of Interferometric Imaging}
\label{sec::Imaging_Background}

\subsection{Interferometric Visibilities}
\label{sec::BasicDefinitions}

Each baseline joining two sites in an interferometer samples complex visibilities. By the van Cittert-Zernike theorem, these visibilities, $V(\mathbf{u})$ are related to the brightness distribution on the sky $I(\mathbf{x})$ via a Fourier transform \citep{TMS}:
\begin{align}
\label{eq::vCZ}
V(\mathbf{u}) &= \int d^2\mathbf{x}\, I(\mathbf{x}) e^{-2\pi i \mathbf{u} \cdot \mathbf{x}},
\end{align}
where $\mathbf{x}$ is an angular coordinate on the sky, in radians, and $\mathbf{u} \equiv \{u, v\}$ is the dimensionless baseline vector, in wavelengths, projected orthogonal to the line of sight.

Interferometry uses a set of measured visibilities $\{ V_i \}$ to estimate the unknown sky image $I(\mathbf{x})$, as we will discuss in \S\ref{sec::InterferometricImaging}. However, when the image is also a function of time, the sampled visibilities at a particular time only represent the corresponding, instantaneous image. In this case, a series of images can be reconstructed if each utilizes only its simultaneous ``snapshot'' visibility coverage. With $N_{\rm s}$ participating sites with mutual visibility of the source, there are at most $N_{\rm s} (N_{\rm s}-1)/2$ visibilities in the snapshot coverage, severely limiting the imaging capabilities when $N_{\rm s}$ is small (see, e.g., Figure~\ref{fig::u-v_Coverage}).

\subsection{Interferometric Imaging via Regularized Minimization}
\label{sec::InterferometricImaging}

We will now review the standard prescription for VLBI imaging via regularized minimization. This prescription encompasses many common approaches to VLBI imaging, such as the maximum entropy method \citep[MEM; see, e.g.,][]{Frieden_1972,Cornwell_Evans_1985,Narayan_Nityananda_1986} and many variants \citep[see, e.g.,][]{Thiebaut_2013,Honma_2014,Lu_2014,Bouman_2016,Chael_2016,Fish_2016_Imaging,Akiyama_2017,Akiyama_2017b} but does not describe iterative deconvolution approaches such as CLEAN \citep{Hogbom_1974}. The flexibility of the regularized minimization framework makes it ideal for sparse and heterogeneous arrays, such as the EHT, and also allows extensions to include, e.g., mitigating the image distortions caused by interstellar scattering \citep{Stochastic_Optics}. 

To simplify our presentation, we will represent reconstructed images $\mathbf{I}$ as square $N {\times} N$ arrays, giving flux density per pixel. 
We denote a sequence of images by $\{ \mathbf{I}_j \}$, where $j$ indexes the time for $N_{\rm t}$ different frames. In the following sections, we will generally treat each image as a vector of length $N^2$ rather than an $N {\times N}$ matrix. Linear operators such as the Fourier transform relating images and interferometric visibilities, blurring via convolution, and discrete gradients are linear and can therefore be represented as $N^2 {\times} N^2$ matrix operators that act on these one-dimensional image vectors (of course, elements of these operators depend on the two-dimensional nature of the images).

Approaches such as MEM estimate the unknown source image $\mathbf{I}$ by numerically minimizing an objective function, $J(\mathbf{I})$. $J$ contains terms that express whether or not an image is consistent with the input VLBI data (a chi-squared term) and also contains terms that favor certain image attributes (such as smoothness or positivity through an entropy or other regularization term). The objective function then takes the form
\begin{align}
\label{eq::Canonical}
J = \chi^2(\mathbf{I},\mathbf{d}) - \alpha_{\rm S} S(\mathbf{I}).
\end{align}
In this expression, $S(\mathbf{I})$ denotes the regularization function for the imaging (e.g., $S(\mathbf{I}) \equiv -\sum_{\ell, m} I_{\ell, m} \ln \left( I_{\ell, m} \right)$ is commonly used for MEM), and $\chi^2$ represents a chi-squared for whatever data products $\mathbf{d}$ are used as part of the imaging. $\alpha_{\rm S}$ is a ``hyperparameter'' that controls the relative weighting of the entropy and data terms. The hyperparameter can be adjusted manually or automatically to yield the expected $\chi^2$ for a satisfactory image \citep[e.g.,][]{Cornwell_Evans_1985} or it can be estimated via cross validation, wherein the data are divided into training and testing sets and the hyperparameters are chosen so that images reconstructed using the training set are compatible with the measurements and errors of the testing set \citep[see][]{Akiyama_2017}. From a probabilistic perspective, the $\chi^2$ term in Eq.~\ref{eq::Canonical} corresponds to a log-likehood while the regularization term corresponds to a log prior distribution of the reconstructed image. 

\subsection{General Prescription for Dynamical Imaging}
\label{sec::DynamicalImaging}
\label{sec::DynamicalImagingFramework}

We now extend this framework and notation to dynamical imaging. In this case, the imaging problem is to simultaneously reconstruct $N_{\rm t}$ different frames $\{ \mathbf{I}_j \}$. Each frame has an associated entropy, and we will average the frame entropies to give a single representative value. Also, the data chi-squared term must be updated so that each data point is compared with its simultaneous reconstructed image. Finally, we will add a new term $\mathcal{R}_x(\{ \mathbf{I}_j \})$ with an associated hyperparameter $\alpha_x$ to regularize the dynamical images (we use $x$ to label different choices for this term). This additional term can enforce expected properties such as continuity from frame to frame, a stable average image, or stable motion. The objective function for dynamical imaging then takes the form
\begin{align}
\label{eq::Objective_Function}
J = \chi^2(\{ \mathbf{I}_j \},\mathbf{d}) - \alpha_{\rm S} \left[ \frac{1}{N_{\rm t}}  \sum_{j=1}^{N_{\rm t}} S(\mathbf{I}_j) \right] + \alpha_x \mathcal{R}_x(\{ \mathbf{I}_j \}).
\end{align}
Note that multiple dynamical regularizers can easily be combined in this framework, and additional regularization terms could be added \citep[e.g., to mitigate interstellar scattering;][]{Stochastic_Optics}. The main purpose of this paper is to develop effective and efficient choices for the dynamical regularization terms $\mathcal{R}_x$ and to test their performance on a variety of simulated data for the EHT.

\subsection{General Considerations for Dynamical Imaging} 
 \label{sec::DynamicalImagingConsiderations}

Before developing specific strategies for dynamical imaging, it is instructive to consider how intrinsic variability can affect image reconstructions that assume a static source. Each baseline changes slowly with the Earth's rotation, so variability of an image on much shorter timescales introduces variations in measured visibilities over small baseline displacements $\Delta \mathbf{u}$. From Eq.~\ref{eq::vCZ}, we see that variations in the visibility over $\Delta \mathbf{u}$ require that the image flux extends over an angular scale $\Delta \mathbf{x} \sim 1/(2\pi |\Delta \mathbf{u}|)$. This mathematical uncertainty relationship arises because the variables of spatial position ($\mathbf{x}$) and spatial wavenumber ($\mathbf{u}$) are Fourier conjugates. One consequence of this property is that, for a static image, variations in the complex visibility seen over a baseline displacement of $\Delta \mathbf{u}$ can be used to infer the image field of view (FOV) without requiring detailed imaging. Likewise, the image field of view determines a maximum averaging time for visibilities sampled from a static image \citep[see \S6.4 of][]{TMS}. 
However, for a variable source interpreted in the context of a static image, rapid intrinsic variability implies the existence of spurious image structure on large scales.  

We now consider some specific examples. First, suppose that variations in the visibility amplitude are seen on a timescale of 5~minutes for an EHT baseline of length $5\,{\rm G}\lambda$. In this case, $|\Delta \mathbf{u}| \sim (5\,{\rm minutes})/(24\,{\rm hours})\times (2\pi) \times (5\,{\rm G}\lambda) \approx 110\,{\rm M}\lambda$. These variations would then imply an image extent of roughly $1/(2\pi |\Delta \mathbf{u}|) \approx 300~\mu{\rm as}$. Note that this inferred extent is an order of magnitude larger than the measured size of \sgra\ at $\lambda=1.3\,{\rm mm}$ \citep[${\approx}40~\mu{\rm as}$;][]{Doeleman_2008}. In addition, the snapshot visibilities can be compared across the exceptionally wide bandwidths of the EHT (4~GHz in 2017, and 18~GHz of spanned bandwidth in 2018 via dual-sideband recording). These also provide $|\Delta \mathbf{u}| \sim (5\,{\rm G}\lambda)\times(4\,{\rm GHz})/(230\,{\rm GHz}) \sim 100\,{\rm M}\lambda$. Thus, visibilities that vary on timescales of minutes but that are stable across the full EHT bandwidth would provide firm evidence of rapid intrinsic variability. 

As another trivial example, no static image can describe data in which the total flux density (i.e., the zero-baseline visibility $V(\mathbf{0})$) is changing with time. The problems of imaging a variable source are further exacerbated with multiple sites because different baseline tracks can cross so that the same spatial Fourier component is sampled at multiple times (see Figure~\ref{fig::u-v_Coverage}).\footnote{For EHT observations of \sgra, SPT-PV and SPT-SMT baselines are very close in u-v space but have a $7.1$ hour offset in sampling. Also, the ALMA-LMT and ALMA-SMT tracks intersect with a time offset of 1.4~hours. See Figure~\ref{fig::u-v_Coverage}.}  

As these examples illustrate, in some cases intrinsic variability can be robustly decoupled from extrinsic sampling variability (from a changing baseline with the Earth's rotation) by constraining the image FOV (effectively imposing an image prior).  In \citet{Lu_2016}, the authors use temporal filtering and normalization of measured visibilities to mitigate intrinsic variability; their chosen filter parameters effectively impose a maximal FOV. However, the strategy of post-processing visibilities has some limitations relative to an image-based approach; for instance, visibility domain smoothing with a baseline-based algorithm does not account for mismatched visibilities on crossing baseline tracks. More generally, visibility-domain averaging of robust observables such as closure phases and closure amplitudes can introduce bias in the measurements. Dynamical imaging addresses both these limitations, providing a framework in which the intrinsic variability is incorporated into the imaging model, so that measurements can be directly compared with reconstructed images without additional averaging.

\section{Regularizers for Dynamical Imaging}
\label{sec::Dynamical_Regularizers}

We now derive three regularizers appropriate for dynamical imaging. Our motivation is to identify regularizers that reflect a range of expected properties for astrophysical cases of interest and that also are efficient to implement in a numerical minimization scheme. Our first regularizer only enforces continuity from frame-to-frame (\S\ref{sec::DiffGaussBlur}), the second favors frames that are small perturbations from the time-averaged image (\S\ref{sec::AvgDiff}), and the third describes an image that evolves approximately as a fluid with a steady motion field (\S\ref{sec::Flow}). We summarize the properties of these regularizers in \S\ref{sec::Regularizer_Summary}.

\subsection{Smoothly Varying Images Over Time}
\label{sec::DiffGaussBlur}

We first develop a generic regularizer that only seeks to enforce continuity from frame to frame in reconstructed images. Because the motion between frames is unknown and may not be constant in time, this regularizer compares the reconstructed flux density of a pixel at one time with the flux density of nearby pixels at a subsequent time. The appropriate definition of ``nearby'' depends on the product of the expected velocity of moving features and the frame interval (which could potentially be irregular). Because this strategy is based on enforcing continuity over short time intervals, we denote the regularizer by $\mathcal{R}_{\Delta t}$. 

Explicitly, we compute the summed difference among all adjacent images after blurring the frames, $\mathbf{I}_j \rightarrow B(\mathbf{I}_j)$, using a circular Gaussian kernel with standard deviation $\sigma_{\Delta t}$. We will focus on two particular choices to define the distance between a pair of images. First, there is the total pixel-by-pixel squared difference:
\begin{align}
\mathcal{D}_{2}(\mathbf{I}, \mathbf{I}') &\equiv \norm{\mathbf{I} - \mathbf{I}'}^2 = \sum_{m, \ell} \left(I_{m,\ell} - I'_{m,\ell}\right)^2.
\end{align}
A simple generalization of this regularizer is to replace the squared norm  $\norm{\dots}^2$ with $\norm{\dots}_p^{p}$ for some fixed $p>0$, $\mathcal{D}_2 \rightarrow \mathcal{D}_p$. 

A second option to define an image distance is the relative entropy (i.e., the Kullback-Leibler divergence): 
\begin{align}
\mathcal{D}_{\rm KL}(\mathbf{I}, \mathbf{I}') &=  \infdiv{\mathbf{I}'}{\mathbf{I}} \equiv \sum_{m, \ell} I'_{m,\ell} \ln\left( \frac{ I'_{m,\ell}}{I_{m,\ell}} \right).
\end{align}
The relative entropy is frequently used to regularize traditional VLBI imaging against a specified image prior for the reconstruction \citep[see, e.g.,][]{Cornwell_Evans_1985} and is also often used for multi-model image registration \citep{Wells_1996,Viola_1997}. Note that the relative entropy is not symmetric, $\mathcal{D}_{\rm KL}(\mathbf{I}, \mathbf{I}') \neq \mathcal{D}_{\rm KL}(\mathbf{I}', \mathbf{I})$, and it need not be positive unless the total flux densities of the two images are equal: $\sum_{\ell,m} I_{\ell, m} = \sum_{\ell,m} I'_{\ell, m}$. Thus, useful alternatives include computing the relative entropy with respect to the normalized images (to preserve positivity of the divergence) and symmetrized versions such as $\frac{1}{2}\left[ \mathcal{D}_{\rm KL}(\mathbf{I}, \mathbf{I}') + \mathcal{D}_{\rm KL}(\mathbf{I}', \mathbf{I}) \right]$ or $\frac{1}{2}\left[ \mathcal{D}_{\rm KL}(\mathbf{I}, \bar{\mathbf{I}}) + \mathcal{D}_{\rm KL}(\mathbf{I}',\bar{\mathbf{I}}) \right]$ with $\bar{\mathbf{I}} \equiv \frac{1}{2}\left( \mathbf{I} + \mathbf{I}' \right)$ (i.e., the Jensen-Shannon divergence). 

The dynamical regularizer then takes the form
\begin{align}
\label{eq::Rdt}
\mathcal{R}_{\Delta t}\left( \left\{ \mathbf{I}_k \right\} \right) &\equiv \sum_{j=1}^{N_{\rm t}-1} \mathcal{D}\left( B\left(\mathbf{I}_j \right), B\left(\mathbf{I}_{j+1} \right) \right).
\end{align}
This regularizer thereby penalizes changes between frames, with steeply decreasing penalty for changes on scales smaller than ${\sim}\sigma_{\Delta t}$. One limitation of the $\mathcal{R}_{\Delta t}$ regularizer is that it does not favor stable ``momentum'' of features between frames. In \S\ref{sec::Flow}, we will discuss an alternative regularizer that favors reconstructions with smooth and stable motion between frames. 
In its simplest implementation, this regularization then depends on only two hyperparameters: $\sigma_{\Delta t}$ and $\alpha_{\Delta t}$ (see \S\ref{sec::DynamicalImagingFramework}). However, note that $\mathcal{R}_{\Delta t}$ is meaningful even in the limit $\sigma_{\Delta t} \rightarrow 0$ (i.e., comparing the total difference between adjacent frames with no blurring applied). This limit is appropriate when the expected motion between consecutive frames is smaller than the finest resolution of reconstructed features (comparable to the nominal array resolution).

This regularization is effective in an imaging framework because the gradient (with respect to changes in each pixel of the $\{ \mathbf{I}_{k} \}$) can be evaluated efficiently. For example, for the $\mathcal{D}_{\rm KL}$ distance function, 
\begin{align}
\label{eq::Rdt_Gradient_KL}
\frac{\partial{\mathcal{R}_{\Delta t}}}{\partial \mathbf{I}_{k}} &= B\left( \left[ \mathbf{1} + \ln \frac{ B(\mathbf{I}_{k})}{B(\mathbf{I}_{k-1})}\right]\delta_{k>1} - \frac{B(\mathbf{I}_{k+1})}{B(\mathbf{I}_{k})} \delta_{k<N_{\rm t}} \right),
\end{align}
where $\mathbf{1}$ denotes a vector of length $N^2$ with every element equal to unity and the indicator function $\delta_x$ is defined to be unity when the subscripted condition $x$ is satisfied and is zero otherwise. Observe that calculating the gradient via Eq.~\ref{eq::Rdt_Gradient_KL} requires roughly $\mathcal{O}\left(N_{\rm t} \times N^2\right)$ computations, while calculating the gradient via finite differences of Eq.~\ref{eq::Rdt} requires roughly $\mathcal{O}\left(N_{\rm t} \times N^4\right)$ computations. Thus, for typically VLBI image reconstructions, which have $N \sim 10^2 - 10^3$, these analytic gradients speed up the imaging by several orders of magnitude. 

Note that in Eq.~\ref{eq::Rdt_Gradient_KL} and throughout this paper, operations such as quotients, powers, norms ($|\dots|$), and products of image vectors are to be computed elementwise. See the Appendix for corresponding expressions for other distance metrics.

\subsection{A Stable Average Image with Small Perturbations}
\label{sec::AvgDiff}

Our next dynamical regularizer is suitable for the case when each snapshot of the time-variable image can be described as a small perturbation from the time-averaged image. This case is applicable for a broad range of stationary processes, such as steady-state accretion or jet systems. Because this regularizer enforces snapshot images to be only small perturbations from the time-averaged image, we denote it $\mathcal{R}_{\Delta I}$.

To proceed, we approximate the time-averaged image by the average of all the reconstructed frames: $\mathbf{I}_{\rm avg} \equiv \frac{1}{N_{\rm t}} \sum_{j=1}^{N_{\rm t}} \mathbf{I}_j$. We then define $\mathcal{R}_{\Delta I}$ to be the summed distance between the estimated time-averaged image and each reconstructed frame:
\begin{align}
\label{eq::RdI}
\mathcal{R}_{\Delta I}\left( \left\{ \mathbf{I}_k \right\} \right)  &= \sum_{j=1}^{N_{\rm t}}  \mathcal{D}\left( \mathbf{I}_{\rm avg}, \mathbf{I}_j \right).
\end{align}
As for $\mathcal{R}_{\Delta t}$, a convenient property of this regularization is that the gradient is efficient to compute (see Appendix). 

Note that this regularization requires only one tunable hyperparameter, $\alpha_{\Delta I}$, determining the overall strength of the regularization. An additional blurring step could be added if individual frames occasionally have flux density in regions that are otherwise empty (e.g., to accommodate flaring behavior), but the average image will tend to act like a blurring operator so we do not expect that this step will normally be needed. Another difference between $\mathcal{R}_{\Delta I}$ and $\mathcal{R}_{\Delta t}$ is that $\mathcal{R}_{\Delta I}$ is insensitive to abrupt changes between frames or even reordering of frames. In this respect, $\mathcal{R}_{\Delta I}$ is analogous to entropy, which is unaffected by the placement of pixels in an image (see \S\ref{sec::InterferometricImaging}).

\subsection{Time-Variable Images with Regular Motion}
\label{sec::Flow}

Our third regularizer is motivated by the case when an image evolves according to a regular prescription for motion -- i.e., a steady flow of flux density over time. In this case, the appearance at one time largely determines the appearance at nearby times. A natural example of this case is an accretion flow, and we will denote this regularization by $\mathcal{R}_{\rm flow}$.  

To proceed, we consider the image $I(x,y,t)$ to be an evolving ``fluid'' with a stable flow vector field $\mathbf{v}(x,y)$. We further assume that the flux density is approximately conserved between nearby frames, so the time-variable images must approximately obey a continuity equation:
\begin{align}
\frac{\partial I(x,y,t)}{\partial t} &= - \nabla \cdot \left[I(x,y,t) \mathbf{v}(x,y)\right]\\
\nonumber &= -\left[ \mathbf{v} \cdot \nabla I + I \nabla \cdot \mathbf{v} \right], 
\end{align}
where $\nabla = \left\{ \partial/\partial x, \partial/\partial y \right\}$ denotes a two-dimensional spatial gradient operator. Hence, at a given time, the image and flow can be combined to estimate the image at a slightly later time:
\begin{align}
\label{eq::Flow_Linear}
\nonumber I(x,y,t + \delta t) &\approx I(x,y,t) + \delta t \frac{\partial I(x,y,t)}{\partial t}\\
&=  I(x,y,t) - \delta t \times ( \mathbf{v} \cdot \nabla I + I \nabla \cdot \mathbf{v} ).
\end{align}

We can now use this approximate forward evolution to regularize multi-frame imaging. We consider a regularizer $\mathcal{R}_{\rm flow}$ that is given by the summed difference between each frame and its predicted values based on linearized forward evolution of the previous frame (via Eq.~\ref{eq::Flow_Linear}) with a discrete spatial gradient operator replacing the continuous gradient. By only comparing adjacent frames, we relax the assumption that the flow field completely determines all forward evolution of a system from an initial state -- we only seek to favor series of images that \emph{approximately} respect a stable flow field over short intervals. For specificity, we will work with the $\mathcal{D}_2$ regularization, in which case,  
\begin{align}
\mathcal{R}_{\rm flow}\left( \left\{ \mathbf{I}_k \right\}, \mathbf{m} \right) &= \sum_{j=1}^{N_{\rm t}-1} \norm[\Big]{ \mathbf{I}_{j+1} - \left( \mathbf{I}_{j} - \nabla \cdot \left[ \mathbf{I}_j \mathbf{m} \right] \right) }^2\\
\nonumber &\hspace{-0.7cm}\approx \sum_{j=1}^{N_{\rm t}-1} \norm[\Big]{\mathbf{I}_{j+1} - \mathbf{I}_{j} + \mathbf{m} \cdot \nabla \mathbf{I}_j + \left(\nabla \cdot \mathbf{m}\right) \mathbf{I}_j }^2\\
\nonumber &\hspace{-0.7cm}\equiv \sum_{j=1}^{N_{\rm t}-1} \norm[\Big]{\mathbf{I}_{j+1} - \lrover{\mathbf{F}}_{\rm flow} \cdot \mathbf{I}_{j}}^2.
\end{align}  
Here, we have replaced the velocity field by a dimensionless motion field $\mathbf{m} = \mathbf{v} \delta t / \delta x$, where $\delta x$ is the discrete grid spacing of reconstructed images, $\delta t$ is the image spacing in time, and $\nabla$ now denotes a finite difference operator that approximates the continuous two-dimensional gradient. We have also defined the linear operator $\lrover{\mathbf{F}}_{\rm flow} \equiv 1 - \mathbf{m} \cdot \nabla - \nabla \cdot \mathbf{m}$, and the second line is an approximation that only becomes exact in the continuous limit because identities such as the product rule do not hold exactly for the discrete gradient operator.\footnote{For example, the one-dimensional finite forward difference operator $\left[ \nabla_{\rm f} x \right]_i \equiv x_{i+1} - x_i$ satisfies $\nabla_{\rm f}(x y) = y \nabla_{\rm f} x + x \nabla_{\rm f} y + (\nabla_{\rm f} x)(\nabla_{\rm f} y)$. Likewise, the analogous finite backward difference operator $\left[ \nabla_{\rm b} x \right]_i \equiv x_{i} - x_{i-1}$ satisfies $\nabla_{\rm b}(x y) = y \nabla_{\rm b} x + x \nabla_{\rm b} y - (\nabla_{\rm b} x)(\nabla_{\rm b} y)$. In the present work, we keep the gradient operator general and assume smooth images with small fractional gradients so that $\nabla(xy) \approx y \nabla x + x \nabla y$.} In this expression and elsewhere, images are treated as one-dimensional vectors, the two-dimensional vector flow is unwrapped to be a one-dimensional vector of 2D motions $m_i = \{ m_{i,x}, m_{i,y} \}$, and products of vectors (e.g., $\left(\nabla \cdot \mathbf{m}\right) \mathbf{I}_j$) are to be computed by multiplying the vectors point-by-point (the Hadamard product). For the construction of this regularizer to be valid (i.e., for the linear approximation of Eq.~\ref{eq::Flow_Linear} to hold), the reconstructed frames must have smooth spatial and temporal gradients; the former is enforced by the image regularization terms $S(\mathbf{I}_j)$, while the latter is enforced by the dynamical regularization. More concretely, the time resolution should be fine enough that the vectors of the motion field do not exceed the nominal VLBI beam that describes the angular resolution of the reconstructed images \citep[analogous to the Courant-Friedrichs-Lewy condition for numerical integration of partial differential equations;][]{CFL_1967}. Thus, observations with finer angular resolution require correspondingly finer temporal resolution. However, this condition does not require observations that are spaced this closely in time; additional frames can be included that do not have corresponding data constraints (see \S\ref{sec::Interpolation}). 

The major difference between the flow regularizer and our previous dynamical regularizers is that, in addition to estimating all the image frames, this reconstruction strategy must simultaneously estimate the flow vector field $\mathbf{m}$. In the Appendix, we provide analytic expressions for the gradients of $\mathcal{R}_{\rm flow}$ with respect to the images and flow, enabling efficient estimation of both in a non-linear minimization framework. 

The motion field can also be regularized (just as the individual frames are regularized) to ensure that it varies smoothly over the image. Because $\mathbf{m}$ is a vector field, one could potentially use the same regularizations as have been proposed for polarimetric synthesis imaging \citep[see, e.g.,][]{Chael_2016,Akiyama_2017b}. However, most of these choices are insensitive to the polarization direction, with the exception of total variation. We will use a closely related choice, the total squared gradient of the velocity field: $\mathcal{R}_{\rm m}(\mathbf{m}) = \norm{\nabla \mathbf{m}_x}^2 + \norm{\nabla \mathbf{m}_y}^2$. This regularizer is also commonly used in studies of optical flow, which reconstruct a flow field from a series of images rather than from sparse Fourier sampling \citep{Horn_Schunck_1981}.\footnote{Another difference between traditional studies of optical flow and our approach is that the former assume an incompressible flow: $\nabla \cdot \mathbf{m} \equiv \mathbf{0}$.} The gradient of $\mathcal{R}_{\rm m}$ with respect to the flow field is simply $\partial \mathcal{R}_{\rm m}/\partial \mathbf{m}_{x,y} = -2\nabla \cdot \left( \nabla \mathbf{m}_{x,y} \right)$. Also, $\mathcal{R}_{\rm m}$ has an associated hyperparameter $\alpha_{m}$ to govern its overall weight.

Alternatively, in some cases the flow may be known or may be adequately modeled with a small number of parameters \citep[see, e.g.,][]{Bouman_2017_StarWarps}. In these cases, dynamical imaging is plausible for much sparser arrays. At the other extreme, with sufficient data, the assumption of a stationary flow can be relaxed and the dynamical imaging could allow a smoothly evolving flow field over time.

\begin{figure}[t]
\centering
\mbox{\large Conventional Imaging}\vspace{-0.3cm}\\
\includegraphics*[clip,width=0.13\columnwidth]{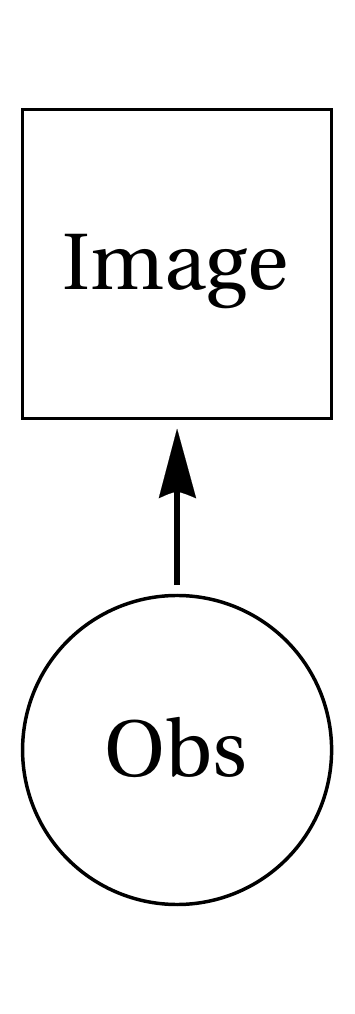}\\\vspace{-0.5cm}\hrulefill\\
\vspace{0.1cm}\mbox{\large Snapshot Imaging}\vspace{-0.15cm}\\
\includegraphics*[clip, trim=0.2cm 0cm 0.2cm 0.0cm,width=0.931\columnwidth]{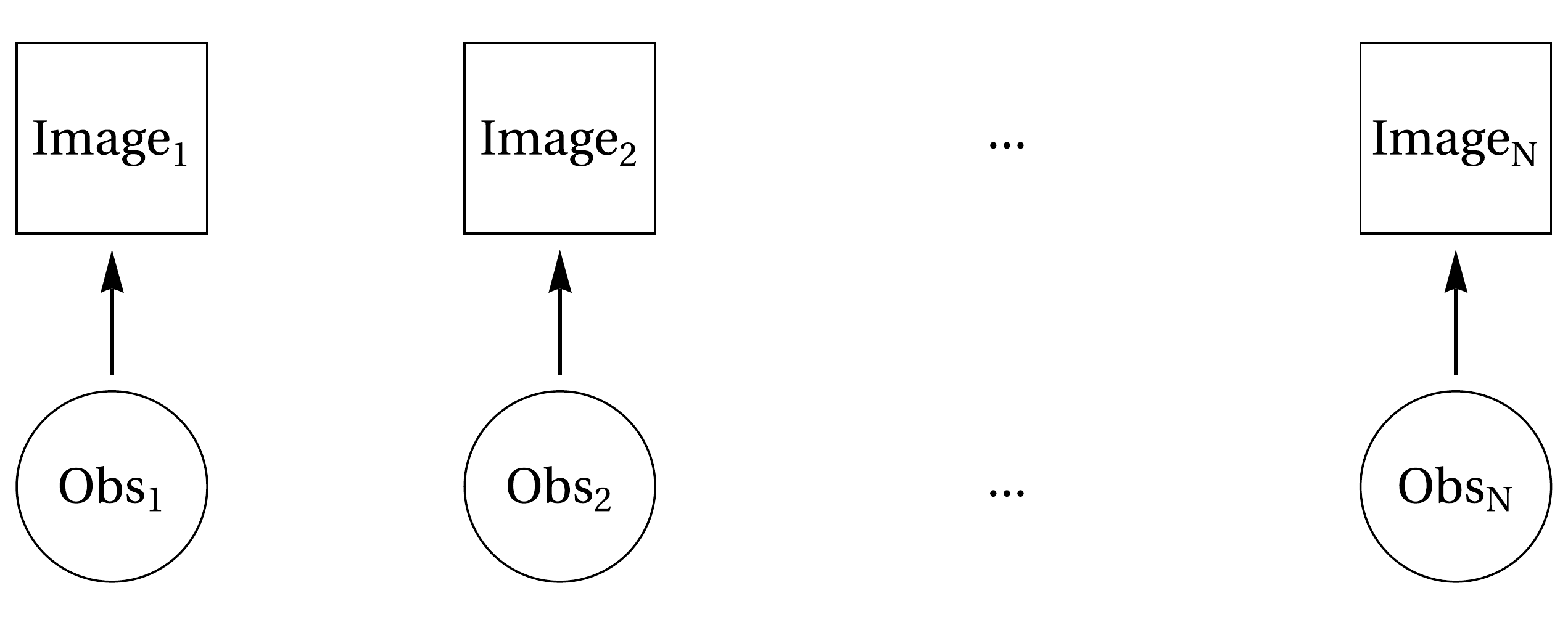}\\\vspace{-0.4cm}\hrulefill\\
\vspace{0.1cm}\mbox{\large Dynamical Imaging: $\mathcal{R}_{\Delta t}$}\vspace{-0.2cm}\\
\includegraphics*[clip, trim=0.2cm 0cm 0.2cm 0.0cm,width=0.931\columnwidth]{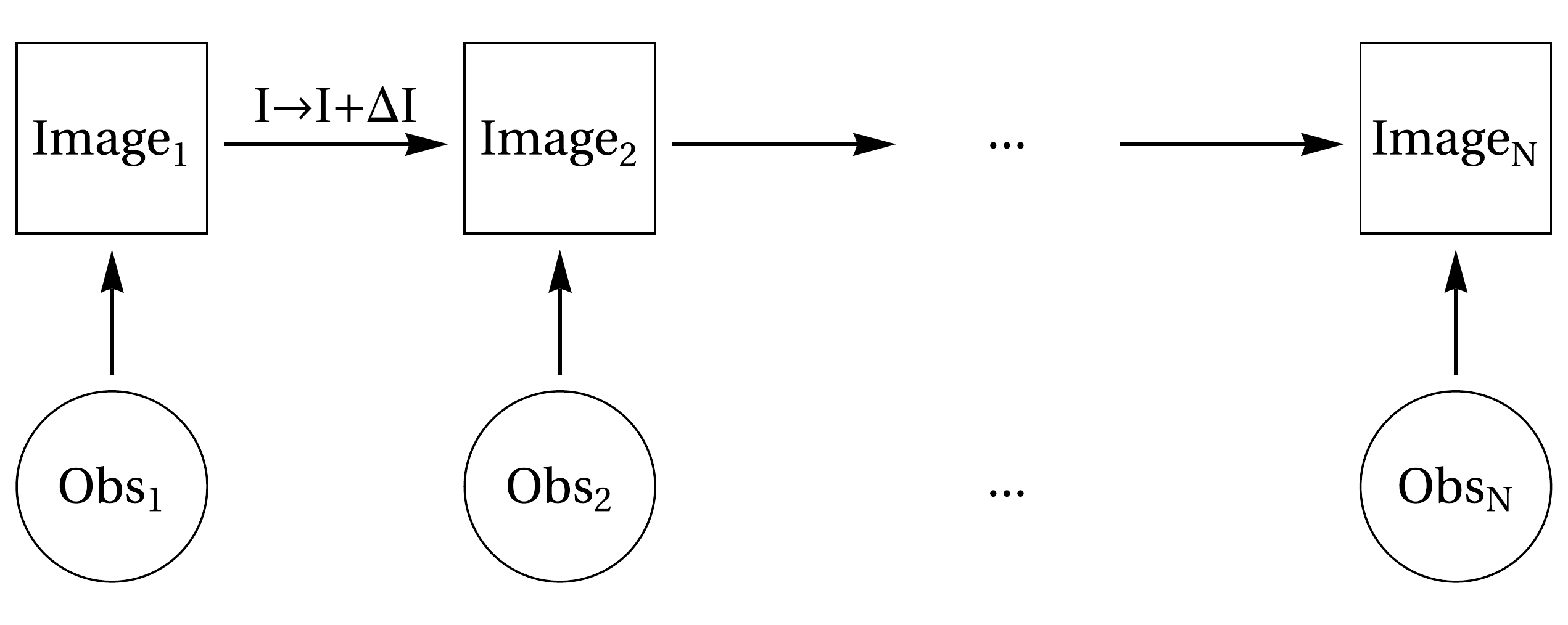}\\\vspace{-0.4cm}\hrulefill\\
\vspace{0.1cm}\mbox{\large Dynamical Imaging: $\mathcal{R}_{\Delta I}$}\vspace{-0.3cm}\\
\includegraphics*[                                 width=0.931\columnwidth]{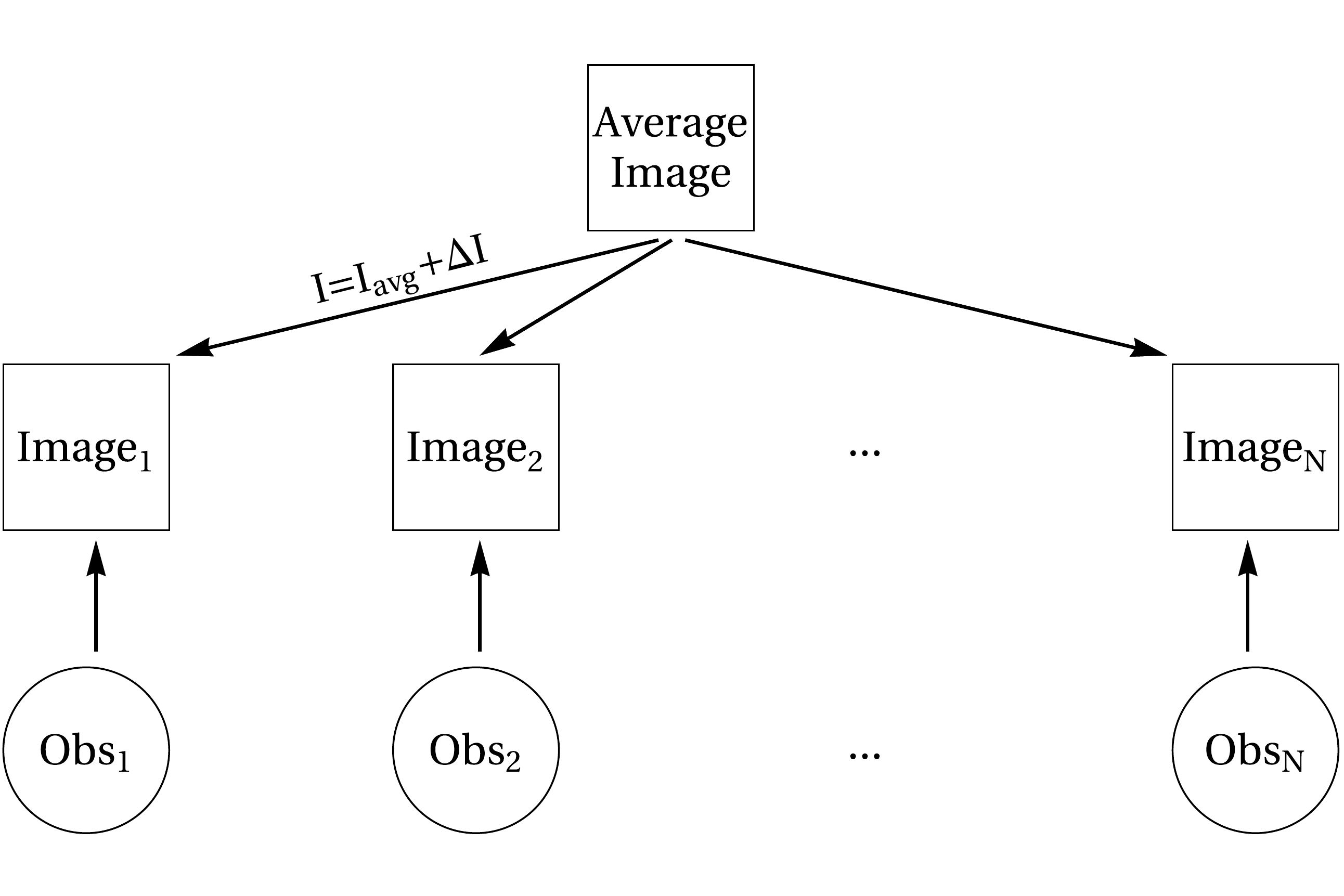}\\\vspace{-0.55cm}\hrulefill\\
\vspace{0.1cm}\mbox{\large Dynamical Imaging: $\mathcal{R}_{\rm flow}$}\vspace{-0.2cm}\\
\includegraphics*[clip, trim=0.2cm 0cm 0.2cm 0.0cm,width=0.931\columnwidth]{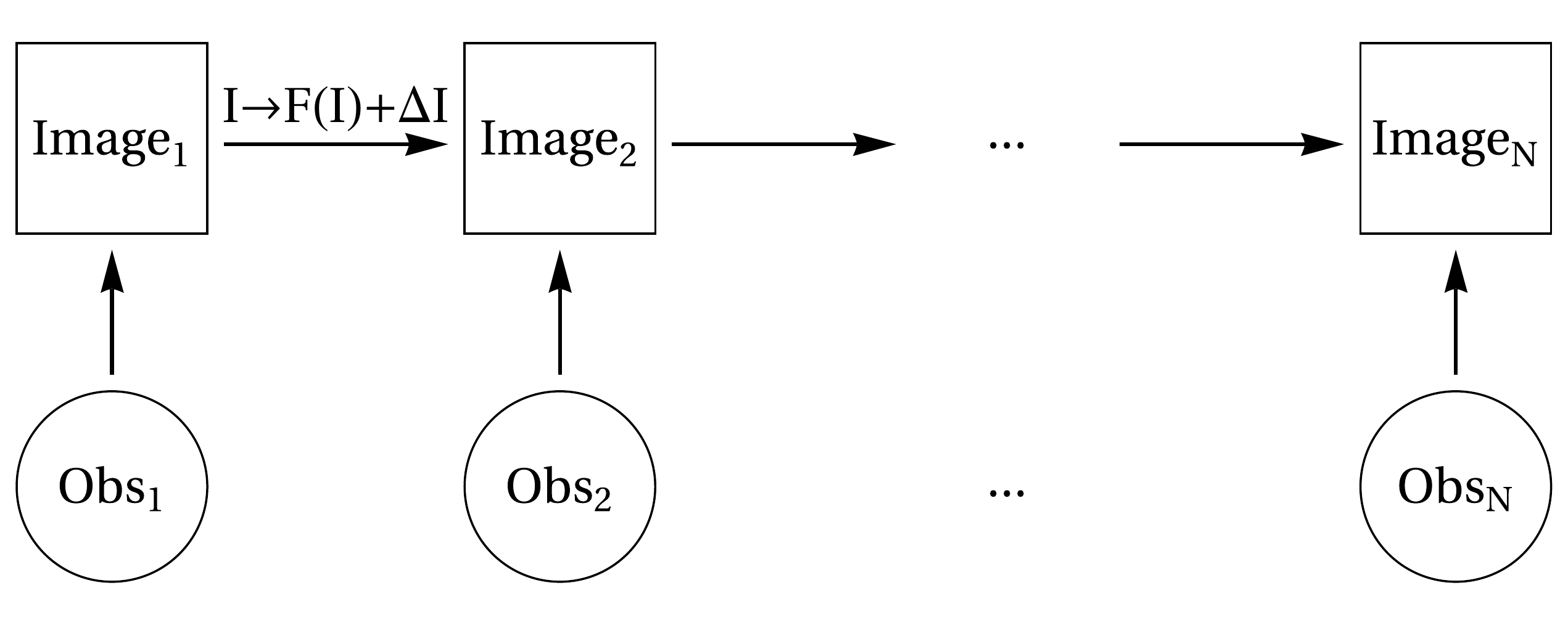}\vspace{-0.19cm}
\caption
{ 
Schematic comparison of our proposed imaging methods. In conventional imaging, a single image is reconstructed from an observation. In snapshot imaging, a set of images is reconstructed from a corresponding set of observations, and each reconstruction is performed independently. For dynamical imaging with $\mathcal{R}_{\Delta t}$ regularization, the images are assumed to be temporally connected, each being a small perturbation of the previous frame. With $\mathcal{R}_{\Delta I}$, each frame is assumed to be a small perturbation of the time-averaged reconstructed frames, and image order is irrelevant. For $\mathcal{R}_{\rm flow}$, each image is a small perturbation of the previous image after forward evolution with the stationary flow, which must be reconstructed along with the images. For the dynamical reconstructions, images can be meaningfully reconstructed even at times with no corresponding observation (see \S\ref{sec::Interpolation}). See \citet{Bouman_2017_StarWarps} for a discussion relating these schematic diagrams to a probabilistic graphical model for the dynamical imaging problem.}
\label{fig::Dynamical_Imaging_Schematic}
\end{figure}

\subsection{Summary and Asymptotic Properties of Dynamical Regularizers}
\label{sec::Regularizer_Summary}

We have developed three regularizers that are suitable for dynamical imaging: $\mathcal{R}_{\Delta t}$, $\mathcal{R}_{\Delta I}$, and $\mathcal{R}_{\rm flow}$. $\mathcal{R}_{\Delta t}$ favors continuity from frame-to-frame within a spatial displacement tolerance determined by $\sigma_{\Delta t}$, $\mathcal{R}_{\Delta I}$ favors frames that are small perturbations from the time-averaged image, and $\mathcal{R}_{\rm flow}$ favors frames that evolve approximately according to a time-independent flow vector field, $\mathbf{m}$ (see Figure~\ref{fig::Dynamical_Imaging_Schematic} for a schematic comparison of these strategies). Each regularizer requires one associated hyperparameter, $\alpha_x$, that assigns overall weight to the dynamical regularization. $\mathcal{R}_{\Delta t}$ also requires one parameter describing the expected angular motion of features from frame-to-frame, and $\mathcal{R}_{\rm flow}$ requires a hyperparameter $\alpha_m$ to regularize the estimated flow field. $\mathcal{R}_{\Delta I}$ requires no additional hyperparameters. These hyperparameters can be fixed according to a priori expectations, they can be treated as Lagrange multipliers and varied to give properties such as a final reduced chi-squared of unity, or they can be estimated using cross validation \citep{Akiyama_2017}. As $\alpha_x \rightarrow 0$, each regularization strategy is equivalent to independently imaging a series of frames. Taking $\alpha_{\Delta t} \rightarrow \infty$ or $\alpha_{\Delta I} \rightarrow \infty$ would enforce a static reconstructed image, equivalent to conventional imaging, although this is not necessarily true for $\alpha_{\rm flow} \rightarrow \infty$. 

It is also possible to normalize each regularizer such that its value is unaffected by the choice of temporal ($\Delta t$; the frame spacing) and spatial resolution ($\Delta x$; the pixel linear dimension) of the reconstruction. As these become arbitrarily small, the dynamical reconstruction approaches a continuous representation in time and space. In particular, the limit $\Delta t \rightarrow 0$ is relevant when including interpolating frames (see \S\ref{sec::Interpolation}), which enable arbitrary temporal resolution. 
The required normalization factor depends on the chosen distance metric. For the $\mathcal{D}_p$ distance, each regularizer $\mathcal{R}_x$ must be multiplied by $\Delta x^{-2(p-1)} \Delta t^{-(p-1)}$. For the Kullback-Leibler divergence $\mathcal{D}_{\rm KL}$ and its symmetrized variants, the normalization factor is $\Delta t^{-1}$. After applying this factor, the associated hyperparameters $\alpha_x$ will be unaffected by the choice of temporal or angular resolution, assuming that the motion in each is well resolved.

\section{Dynamical Imaging and Interpolation}
\label{sec::Interpolation}

Dynamical imaging also serves as a framework for temporal interpolation of images. Namely, image frames can be added, even at times when there are no corresponding data. Without dynamical regularization, these additional frames would default to the image that maximizes the entropy (typically an image with constant brightness, possibly uniformly zero). However, dynamical imaging will favor frames that respect the chosen regularizer. For example, when using $\mathcal{R}_{\Delta t}$, the additional frames will converge toward images that enforce continuity of features with the nearest data-constrained frames. For $\mathcal{R}_{\Delta I}$, frames without data will default to the estimated time-averaged image. For $\mathcal{R}_{\rm flow}$, unconstrained frames will interpolate according to the derived flow map. In each case, frames with missing data can inherit partial information from other times. Moreover, for the case of $\mathcal{R}_{\rm flow}$, frames can be intentionally spaced at finer resolution than the sampling time to ensure that the linear approximation of Eq.~\ref{eq::Flow_Linear} is accurate. All these strategies will produce different results than a straightforward linear interpolation between images, as is commonly used to visualize multi-epoch VLBI studies \citep[e.g.,][]{Lister_2016}. 

However, we have found that the interpolated frames can sometimes have a different appearance than data-constrained frames. For instance, when using the $\mathcal{R}_{\Delta t}$ regularization with $\sigma_{\Delta t} > 0$, the interpolated frames are ``blurred out'' relative to the data-constrained frames. This blurring is unsurprising, as it helps to minimize the mean-squared difference among adjacent frames as elements of flux move in time. Consequently, the interpolated frames may achieve continuity of features but may have temporal discontinuities in the total flux density or image entropy. 

To mitigate the artifacts in interpolated frames, we can directly enforce continuity of quantities such as flux density and entropy. To do so, we add corresponding terms to the objective function (Eq.~\ref{eq::Objective_Function}). For example, to enforce continuity of image entropy, one can add,
\begin{align}
\mathcal{R}_{\Delta S} &\equiv \sum_{j=1}^{N_{\rm t}-1} \left[ S(\mathbf{I}_j) - S(\mathbf{I}_{j+1}) \right]^2,
\end{align}
weighted by an associated hyperparameter $\alpha_{\Delta S}$. The hyperparameter can be adjusted so that this term allows continuous variations of the entropy among frames without forcing the entropy of each frame to be equal. 
The gradient of this term is straightforward to compute in terms of the single-image gradients:
\begin{align}
\frac{ \partial \mathcal{R}_{\Delta S}}{\partial \mathbf{I}_k} &= 2 \Big\{\left[ S(\mathbf{I}_j) - S(\mathbf{I}_{j-1}) \right]\delta_{k>1} \\
\nonumber & \qquad {} + \left[ S(\mathbf{I}_j) - S(\mathbf{I}_{j+1}) \right]\delta_{k<N_{\rm t}}\Big\} \frac{ \partial S(\mathbf{I}_k)}{\partial \mathbf{I}_k}.
\end{align}
Likewise, to make the total flux continuous from frame to frame, we can add
\begin{align}
\mathcal{R}_{\Delta F} &\equiv \sum_{j=1}^{N_{\rm t}-1} \left[ F(\mathbf{I}_j) - F(\mathbf{I}_{j+1}) \right]^2,
\end{align}
weighted by an associated hyperparameter $\alpha_{\Delta F}$, where $F(\mathbf{I}) \equiv \sum_{\ell,m} I_{\ell,m}$ denotes the total flux density of an image. The gradient is again straightforward to compute:
\begin{align}
\frac{ \partial \mathcal{R}_{\Delta F}}{\partial \mathbf{I}_k} &= 2 \Big\{ \left[ F(\mathbf{I}_j) - F(\mathbf{I}_{j-1}) \right]\delta_{k>1} \\
\nonumber & \qquad  {} + \left[ F(\mathbf{I}_j) - F(\mathbf{I}_{j+1}) \right]\delta_{k<N_{\rm t}}\Big\} \frac{ \partial F(\mathbf{I}_k)}{\partial \mathbf{I}_k},
\end{align}
where $\partial F(\mathbf{I}_k)/\partial \mathbf{I}_k$ is a vector with each element equal to $1$ and of length equal to the number of pixels in image $\mathbf{I}_k$.

{
\begin{deluxetable}{lc}
\tablewidth{0.9\columnwidth}
\tablecaption{Assumed Site System Equivalent Flux Densities (SEFD).}
\tablehead{ 
\colhead{ Site } & \colhead{ SEFD (Jy) }}   
\startdata
SMA/JCMT      & 4900\\
SMT      & 11900\\
LMT      & 560\\
ALMA/APEX     & 220\\
SPT      & 1600\\
PdB      & 1600\\
PV       & 2900\\[0.1em]
\tableline \\[-0.7em]
CA     & 10000\\
KP       & 10000
\enddata
\label{tab::EHT}
\tablecomments{ Most SEFDs match what was specified in the 2016 EHT call for proposals. CA and KP are as-yet hypothetical EHT sites at the location of the CARMA array and at Kitt Peak, respectively.}
\end{deluxetable}
}

\section{Examples of Dynamical Imaging}
\label{sec::Imaging_Examples}

We will now show a few representative examples of dynamical imaging using simulated VLBI observations, and we will discuss general trends that we have identified. We conclude this section with an example showing frames from dynamical imaging of M87.

\subsection{Implementation and Procedure}

We implemented the dynamical regularizers developed in \S\ref{sec::Dynamical_Regularizers} as an extension to the \texttt{eht-imaging}\footnote{\url{https://github.com/achael/eht-imaging}} Python library, which was originally developed for polarimetric VLBI imaging \citep[][]{Chael_2016}. This library provides a modular and flexible imaging framework that can utilize a variety of imaging regularizers (e.g., entropy, total variation, and $\ell_p$) and arbitrary combinations of data constraints (e.g., complex visibilities, the bispectrum, or closure quantities). 
We also used this library for generating synthetic data. Except when noted otherwise, we chose observing parameters that correspond to the 2017 EHT: an observing bandwidth of 4\,GHz and site system equivalent flux densities given in Table~\ref{tab::EHT}. For simplicity, our simulated observations of \sgra\ account for sensitivity losses from the blurring effects of interstellar scattering \citep{Fish_2014} but not irregular, refractive effects \citep{Johnson_Gwinn_2015}.  

The fundamental interferometric data product is the sampled complex interferometric visibility (Eq.~\ref{eq::vCZ}). However, because of a large stochastic contribution from the atmosphere to each site's phase, high-frequency VLBI arrays can typically only measure quantities such as closure phase robustly \citep{TMS}. Imaging algorithms can then work with these robust data products directly \citep[see, e.g.,][]{Buscher_1994,Baron_2010,Chael_2016,Bouman_2016,Akiyama_2017}. Nevertheless, in the near future, improved techniques such as simultaneous subarrayed observations of calibrators \citep[see][]{Broderick_2011} may provide absolute phase information. Also, we expect dynamical imaging to be applicable at the lower frequencies where phase referencing is routine; e.g., observations with the VLBA at wavelengths of $3\,{\rm mm}$ and longer. Thus, we will show results both when using complex visibilities and when using only visibility amplitudes and closure phase. 
 
To minimize the objective function given by Eq.~\ref{eq::Objective_Function} (i.e., to perform dynamical imaging), we used the non-linear minimization package {\tt optimize.minimize} of {\tt SciPy} \citep{SciPy}. We used the Limited-Memory BFGS algorithm (L-BFGS) \citep{Byrd_1995} except when memory requirements to store the partial Hessian became prohibitive (generally when imaging $\gg$100 frames simultaneously), in which case we instead used the conjugate gradient algorithm implemented in {\tt SciPy} (which does not compute the Hessian). 

Similar to conventional VLBI imaging, convergence to the minimum of the objective function for dynamical imaging can be challenging because of the extremely high-dimensional ($N^2 \times N_{\rm t} \gsim 10^6$) parameter space surveyed. Convergence is especially challenging when using only robust VLBI observables, such as closure phases, rather than complex visibilities because the relative image centroid among frames is only constrained by the dynamical regularization. Consequently, we used a number of strategies to assist convergence, most involving multiple iterations of minimization with modified initial values. One particularly effective strategy for avoiding local minima, following \citet{Chael_2016}, is to repeatedly image the data, re-initializing the minimization each time to be equal to the previous reconstructed images convolved with the nominal VLBI array resolution. In cases with many high-quality data points for each snapshot, we iterated between imaging all frames and allowing convergence to proceed on individual frames independently. For $\mathcal{R}_{\Delta I}$, we repeatedly re-initialized all frames to the current average image. For $\mathcal{R}_{\rm flow}$, we repeatedly re-initialized the flow to be uniformly zero. In all cases, we determined the dynamical imaging hyperparameters $\alpha_x$ by making them as large as possible while still achieving a final reduced $\chi^2$ near unity.  

One modification to the prescription outlined above that we did find to be effective for larger arrays, such as the VLBA, was to apply the dynamical regularizers to the logarithm of the reconstructed images rather than to the images when using the $\mathcal{D}_2$ or $\mathcal{D}_p$ distance function (here, we assume image positivity). This change helps the dynamical regularization to improve time variable imaging of faint image features and significantly improved reconstructed images with dynamic range $\gsim 100$, although it is unnecessary when using the relative entropy distance function $\mathcal{D}_{\rm KL}$. 

To assess the fidelity of reconstructed images when the true (model) image is known, we utilize the (normalized) mean squared error (MSE):
\begin{align}
{\rm MSE} \equiv \left. \left[\sum_{\ell,m} \left(I_{\ell,m} - I_{\ell,m}'\right)^2 \right]\middle/\left[\sum_{\ell,m} I_{\ell,m}^2 \right]\right..
\end{align}
Here, $I_{\ell,m}$ is the model image and $I_{\ell,m}'$ is the reconstructed image. In cases where closure phases are used for image reconstructions, the image centroid is unconstrained and we report the MSE that is minimum over all shifts $\{ \Delta \ell, \Delta m\}$. The precise value of the MSE should not be taken too literally because it is sensitive to sharp features in the original image that the array cannot resolve \citep[see also][]{Gomes_2017}. Nevertheless, the MSE does tend to provide a crude characterization of reconstructed image quality.

\begin{figure*}[t]
\centering
\includegraphics*[width=0.86\textwidth]{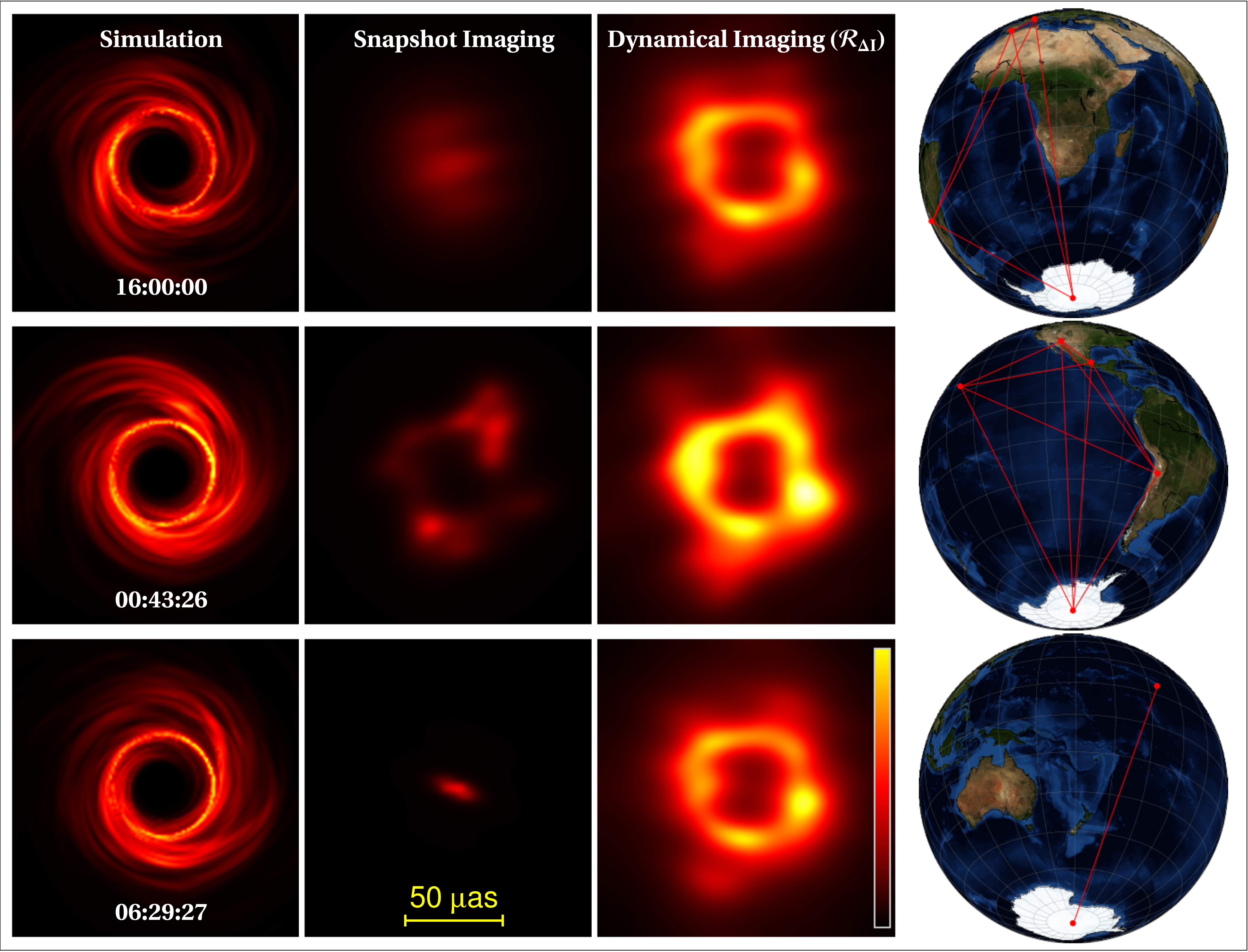}
\caption
{ 
Example reconstruction of a face-on accretion disk with and without dynamical regularization. The full reconstruction comprised 100 frames of 8.8~minutes each beginning at 16:00~GST and spanning a total of 14.6~hours. 
The simulated images are from a 3D GRMHD simulation \citep[\texttt{b0-high} from][]{Shiokawa_2013}. The above panels show the simulated images, snapshot reconstructions (using conventional maximum entropy imaging with only the instantaneous u-v coverage), dynamical reconstructions (using $\mathcal{R}_{\Delta I}$), and the baseline coverage at three times. The color scale is linear and is consistent among different times but is scaled separately for each case based on the maximum brightness over all frames. 
Because the early and late frames have few data constraints, the snapshot image reconstructions of those frames are almost entirely uninformative and poorly approximate the true images. In contrast, the dynamical imaging reconstructions at those times appear almost identical, with the data only supporting small perturbations from the estimated time-averaged image. 
}
\label{fig::RdI_Example}
\end{figure*}

\begin{figure}[t]
\centering
\includegraphics*[width=\columnwidth]{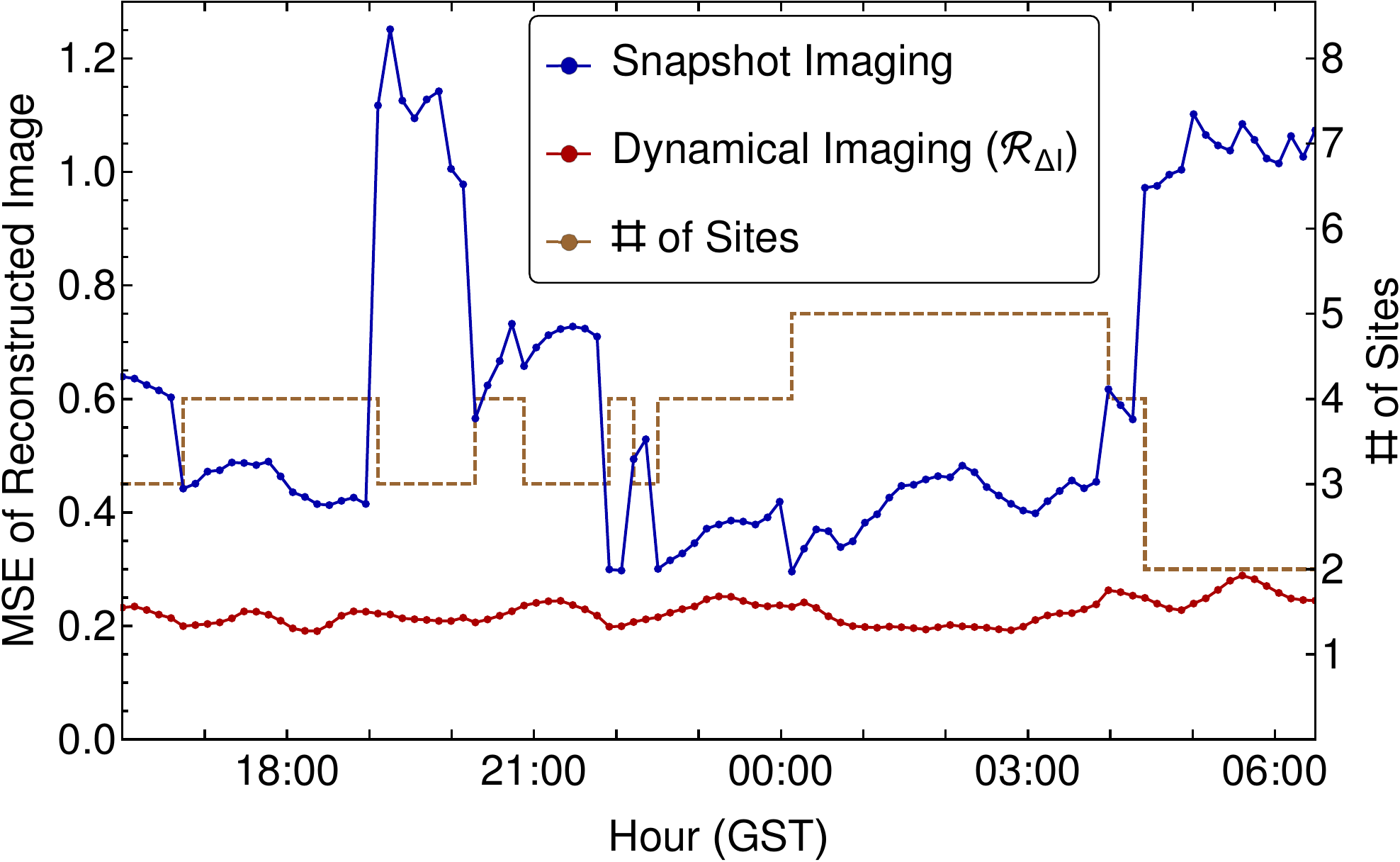}
\caption
{ 
Mean-squared error (MSE) as a function of time for all frames of the reconstruction shown in Figure~\ref{fig::RdI_Example} when compared with the simultaneous simulated frames. As expected, the snapshot reconstructions vary erratically with the most abrupt changes occurring when \sgra\ rises or sets at a participating site (here, we use an elevation limit of $15^\circ$; see Figure~\ref{fig::u-v_Coverage}). In contrast, the MSE of reconstructions with dynamical imaging are relatively steady, demonstrating the added resilience of snapshot reconstructions when using all data concurrently. Nevertheless, with this sparse array, most of the improvement in MSE comes from the superior estimate of the time-averaged image rather than from precisely tracking the changing features of the image. 
}
\label{fig::RdI_Example_MSE}
\end{figure}

\begin{figure}[t]
\centering 
\vspace{1.4cm}
\includegraphics*[width=\columnwidth]{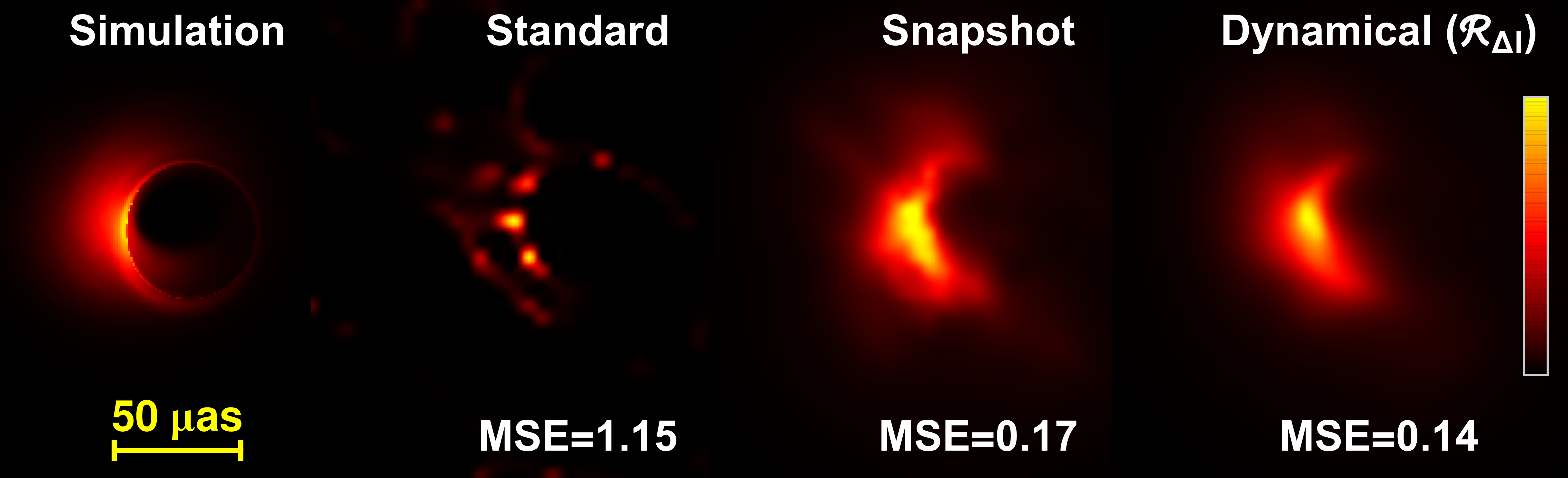}
\caption
{ 
Comparing the time-averaged image of a simulated accretion flow (left) to three reconstruction strategies: standard VLBI imaging that assumes a static source (left-center), time-averaged snapshot image reconstructions (right-center), and time-averaged dynamical imaging with $\mathcal{R}_{\Delta I}$ regularization (right). For the standard and snapshot imaging, we used maximum entropy imaging. 
In this example, we matched the simulation and observing parameters given in \citet{Lu_2016} (here, a single 12-hour observation). As expected, conventional VLBI imaging works poorly, but both averaged snapshot imaging and dynamical imaging are comparable in quality to scaling, averaging, and smoothing the interferometric visibilities before static imaging \citep[$\mathrm{MSE}{=}0.14$;][]{Lu_2016}. Here and throughout this paper, the color scale is linear.\vspace{0.9cm}
}
\label{fig::TimeAveragedImage}
\end{figure}

\subsection{Dynamical Imaging of a Steady Accretion Flow}

To examine the potential capabilities of dynamical imaging with EHT data, we generated synthetic data from a face-on view of a 3D GRMHD simulation of an accretion flow onto \sgra\ \citep[\texttt{b0-high} from][]{Shiokawa_2013}. Figure~\ref{fig::RdI_Example} shows image reconstructions using both snapshot imaging (see Figure~\ref{fig::Dynamical_Imaging_Schematic}) and dynamical imaging. These reconstructions using the $\mathcal{R}_{\Delta I}$ regularizer with the $\mathcal{D}_2$ distance metric and complex visibilities for the data product; we reconstructed 99 frames, each $24 t_{\rm G} \approx 530~{\rm seconds}$, for a total duration of 14.6~hours. At times with poor u-v coverage, the snapshot images are uninformative while frames from dynamical imaging are close to the estimated time-averaged image. Dynamical imaging successfully identifies the time-variable regions of enhanced flux density and can also identify the flow direction, which is not apparent in the snapshot reconstructions. 

To quantify the improvement of dynamical imaging relative to snapshot imaging, Figure~\ref{fig::RdI_Example_MSE} shows the MSE as a function of time for these two reconstructions. Notably, the MSE for snapshot imaging changes significantly over the observation, increasing steeply when the number of sites with mutual visibility of \sgra\ drops. In contrast, the MSE for dynamical imaging is lower overall and is steady, showing the increased resilience to limited data.

\subsection{Estimates of Time-Averaged Images}

Another important utility of dynamical imaging is to estimate the time-averaged image over an observation. For studies of \sgra\ with the EHT, the time-averaged image is of intense interest because it may reveal distinctive features of the spacetime near a black hole such as the black hole ``shadow'' \citep{Bardeen_1972,Luminet_1979,Falcke_2000,Takahashi_2004,Johannsen_2010}. Yet, as discussed in \S\ref{sec::DynamicalImagingConsiderations}, imaging techniques that assume a static source can be severely affected by intrinsic source variability; conventional imaging will \underline{not} simply provide an estimate of the time-averaged image. Instead, the variability must be integrated into the imaging procedure by either preprocessing the data to render it compatible with a static source assumption \citep{Lu_2016} or by modifying the imaging procedure to accommodate image variability, as we propose here.

To test this application of dynamical imaging, we used the simulated EHT data from \citet{Lu_2016} for a GRMHD simulation of an accretion flow onto \sgra. Note that, in contrast with our other examples, this dataset included the CARMA array (the CARMA observatory was shut down in 2015), it sampled the images with 16 GHz of bandwidth rather than 4 GHz, and it used slightly different SEFDs than are given in Table~\ref{tab::EHT}. Because the frame spacing is rather large in this example (3.7 minutes), we again used the regularizer $\mathcal{R}_{\Delta I}$ with the $\mathcal{D}_2$ distance metric. 

Figure~\ref{fig::TimeAveragedImage} compares the time-averaged estimates from dynamical imaging with the time-averaged simulated image. The estimated average image is comparable in quality to the image obtained with the scaling, averaging, and smoothing approach of \citet{Lu_2016}. We also found that averaging the snapshot images gives an estimated average image with comparable quality as these more sophisticated approaches, especially if periods with poor u-v coverage were downweighted or omitted. Thus, a weighted average of snapshot images, favoring times with superior u-v coverage, may also produce reliable estimates of the time-averaged image and will provide a useful comparison for these other approaches.

\subsection{Dynamical Imaging of Flares}

\begin{figure*}[h]
\centering
\includegraphics*[clip, trim=0.2cm 0cm 0.2cm 0.0cm,width=\textwidth]{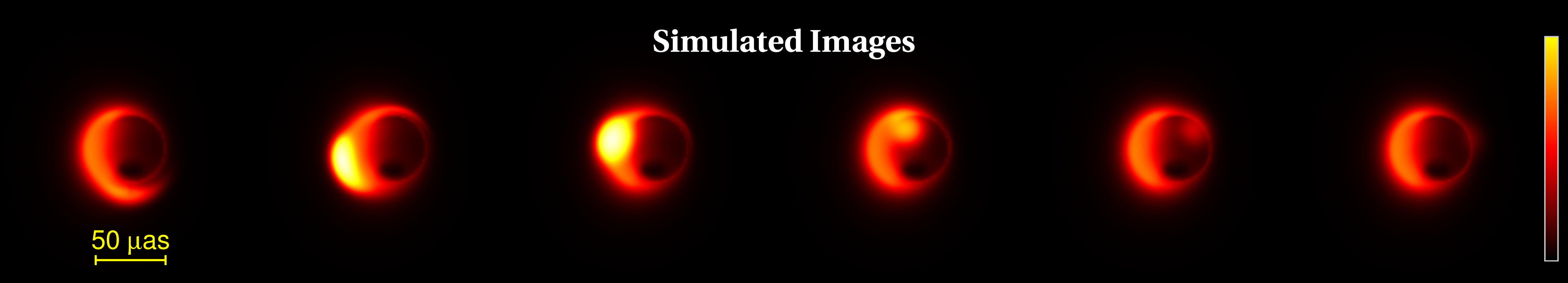}\vspace{-0.2cm}
\includegraphics*[clip, trim=0cm 0.5cm 0cm 5cm,width=\textwidth]{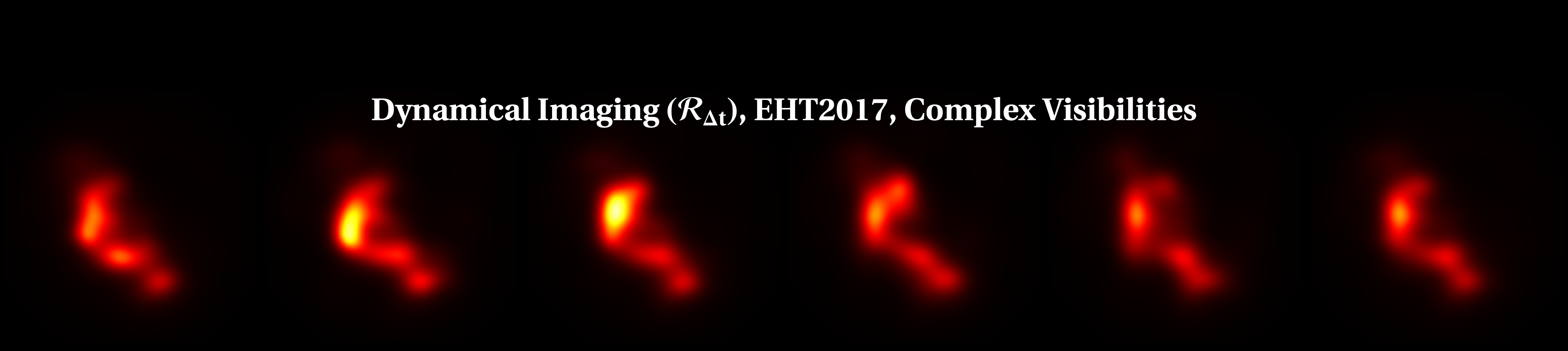}\vspace{-0.2cm}
\includegraphics*[clip, trim=0cm 0.5cm 0cm 5cm,width=\textwidth]{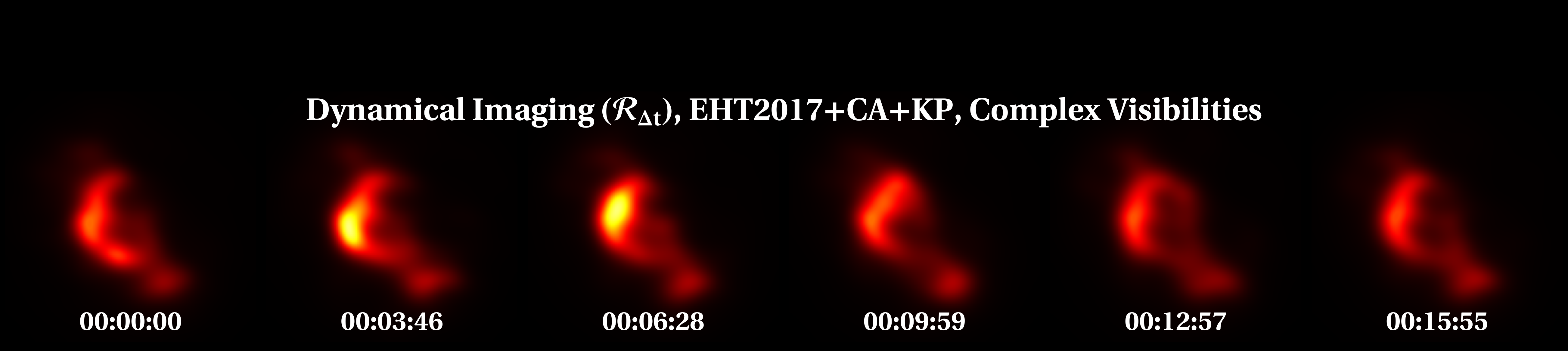}
\caption
{ 
Example dynamical reconstruction of a simulated flare. This simulation is a ``hot spot'' orbiting \sgra\ with a period of 27 minutes \citep[model B of][]{Doeleman_Hotspots}. The total observation covers one full orbit of the hot spot. Top panels show six selected frames of the simulated images, middle panels show corresponding reconstructions with the 2017 EHT array, and bottom panels show reconstructions with the 2017 EHT array plus sites at the location of the CARMA array and at Kitt Peak, each with an assumed system equivalent flux density (SEFD) of $10{,}000~{\rm Jy}$. These reconstructions used $\mathcal{R}_{\Delta t}$ regularization and complex visibilities with only thermal noise added. The spurious structure to the southwest in each reconstruction reflects the significantly anisotropic beam, which contains almost no power at this location during the simulated GST range because of a void in the u-v coverage at the corresponding (orthogonal) position angles (see Figure~\ref{fig::u-v_Coverage}). 
}
\label{fig::Hotspot_Example_ComplexVis}
\end{figure*}

\begin{figure*}[h]
\centering
\includegraphics*[clip, trim=0.2cm 0cm 0.2cm 0.0cm,width=\textwidth]{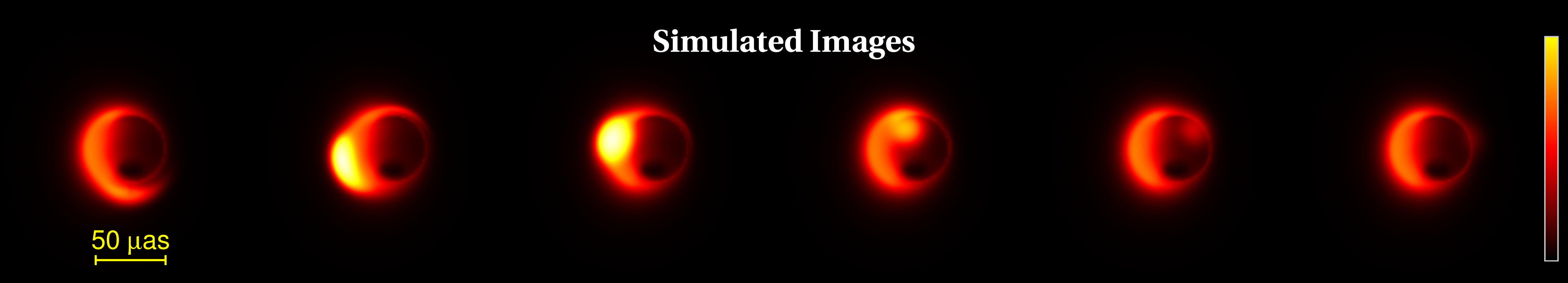}\vspace{-0.2cm}
\includegraphics*[clip, trim=0cm 0.5cm 0cm 5cm,width=\textwidth]{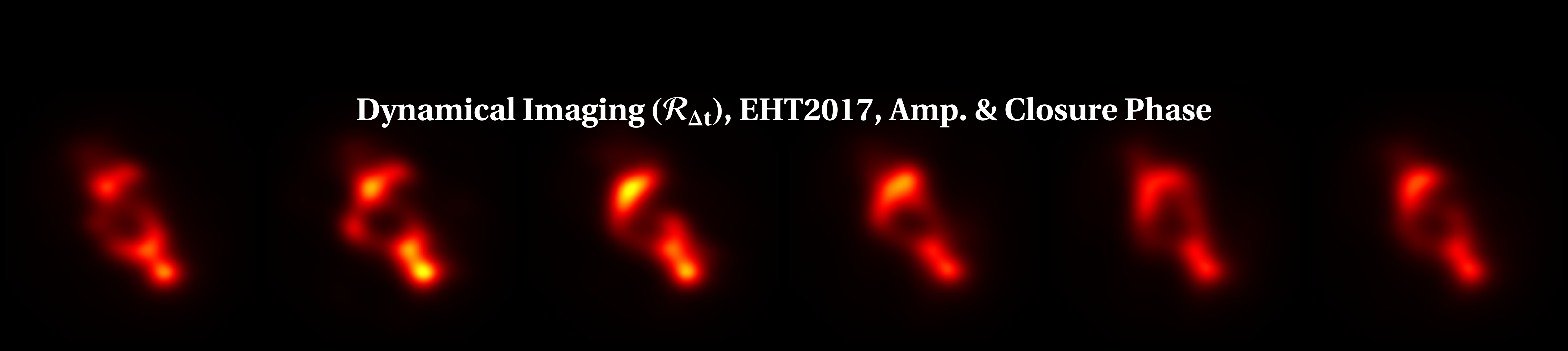}\vspace{-0.2cm}
\includegraphics*[clip, trim=0cm 0.5cm 0cm 5cm,width=\textwidth]{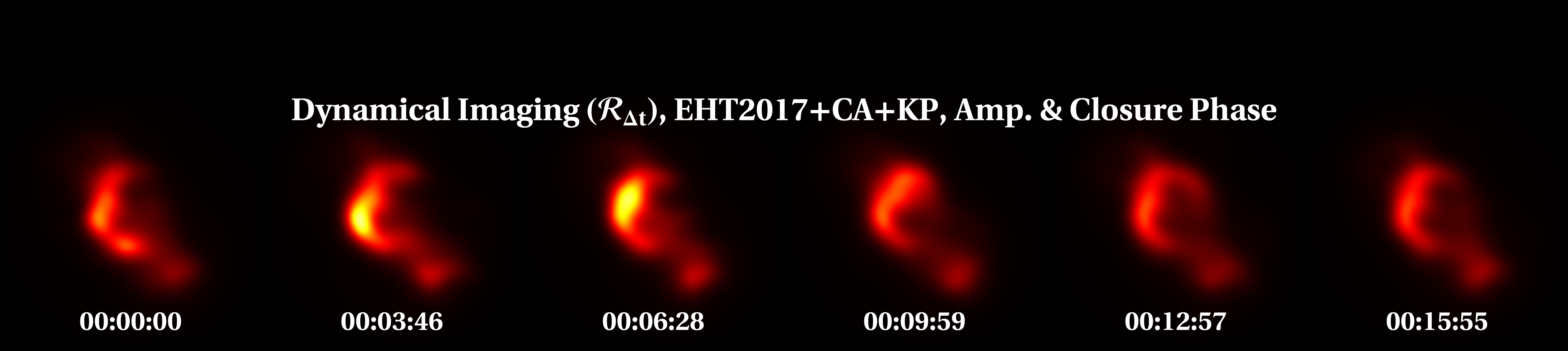}
\caption
{ 
Same as Figure~\ref{fig::Hotspot_Example_ComplexVis} but using only visibility amplitudes and closure phases for the reconstructed images. 
}
\label{fig::Hotspot_Example_ClP_Amp}
\end{figure*}

Another important application for dynamical imaging is to study flares of \sgra\ via direct imaging. Figures~\ref{fig::Hotspot_Example_ComplexVis} and \ref{fig::Hotspot_Example_ClP_Amp} show example reconstructions for an orbiting ``hot spot'' near \sgra\ \citep{Broderick_Loeb_2006}, using simulated observations that span only 27 minutes. For these examples, we used $\mathcal{R}_{\Delta t}$ regularization with the symmetrized KL divergence as the distance metric. While this simulated observation is too short to build up significant baseline coverage via Earth rotation to estimate an accurate time-averaged image, the reconstructions successfully identify the motion of the hot spot, especially if additional sites at Kitt Peak and CARMA are included. Thus, in the coming years, the EHT may be able to trace rapidly evolving structures and estimate orbital rotation curves from dynamical imaging, especially if the array continues to expand.

\begin{figure*}[t]
\centering
\includegraphics*[width=\textwidth]{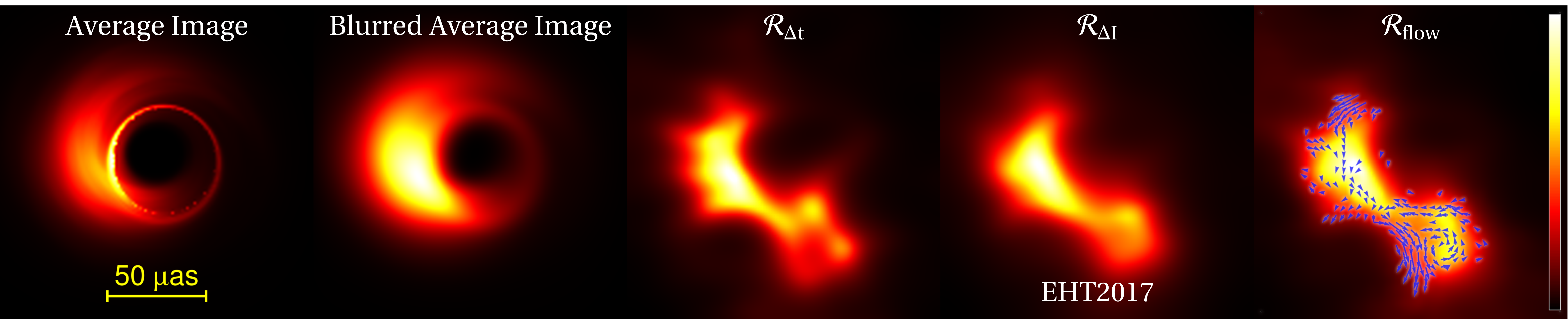}
\caption
{ 
Comparison of dynamical imaging methods for an accretion flow viewed at $30^\circ$ off the rotation axis. Left two panels show the average simulated image and the average simulated image blurred with half the nominal observing array beam (the full beam is $26\,\mu{\rm as} \times 16\,\mu{\rm as}$ with the major axis at a position angle of $73^\circ$ east of north). Remaining panels show the average reconstructed image after dynamical imaging with $\mathcal{R}_{\Delta t}$, $\mathcal{R}_{\Delta I}$, and $\mathcal{R}_{\rm flow}$ regularization, respectively. The flow field that was derived when using $\mathcal{R}_{\rm flow}$ regularization is overplotted on the final image; flow vectors (scaled in length by a factor of 5) show the derived motion from frame-to-frame (i.e., over an interval of 67 seconds). As this example illustrates, EHT coverage in 2017-2018 is unlikely to be sufficient to derive a reliable flow field without additional constraints on the flow structure. Nevertheless, the reconstructed images with $\mathcal{R}_{\rm flow}$ are broadly consistent with the other methods.  
}
\label{fig::Method_Compare}
\end{figure*}

\begin{figure*}[t]
\centering
\includegraphics*[width=\textwidth]{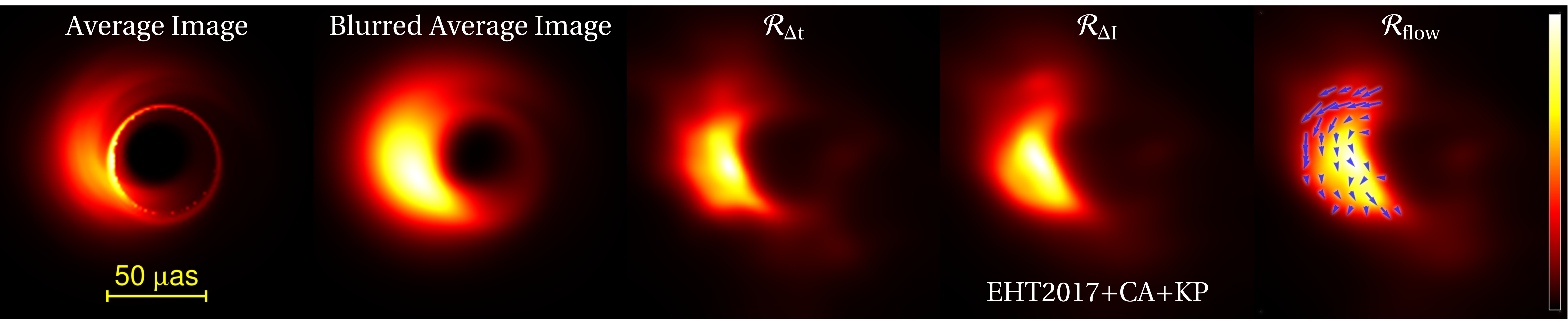}
\caption
{ 
Same as Figure~\ref{fig::Method_Compare}, but using the 2017 EHT array plus sites at the location of the CARMA array and at Kitt Peak. In this figure, the derived motion field vectors are scaled by a factor of 10. While the two additional sites hardly change the observing beam (the blurred image is nearly identical to Figure~\ref{fig::Method_Compare}), they significantly improve the snapshot baseline coverage and the dynamical imaging. 
}
\label{fig::Method_Compare_2}
\end{figure*}

\begin{figure}[t]
\centering
\includegraphics*[width=\columnwidth]{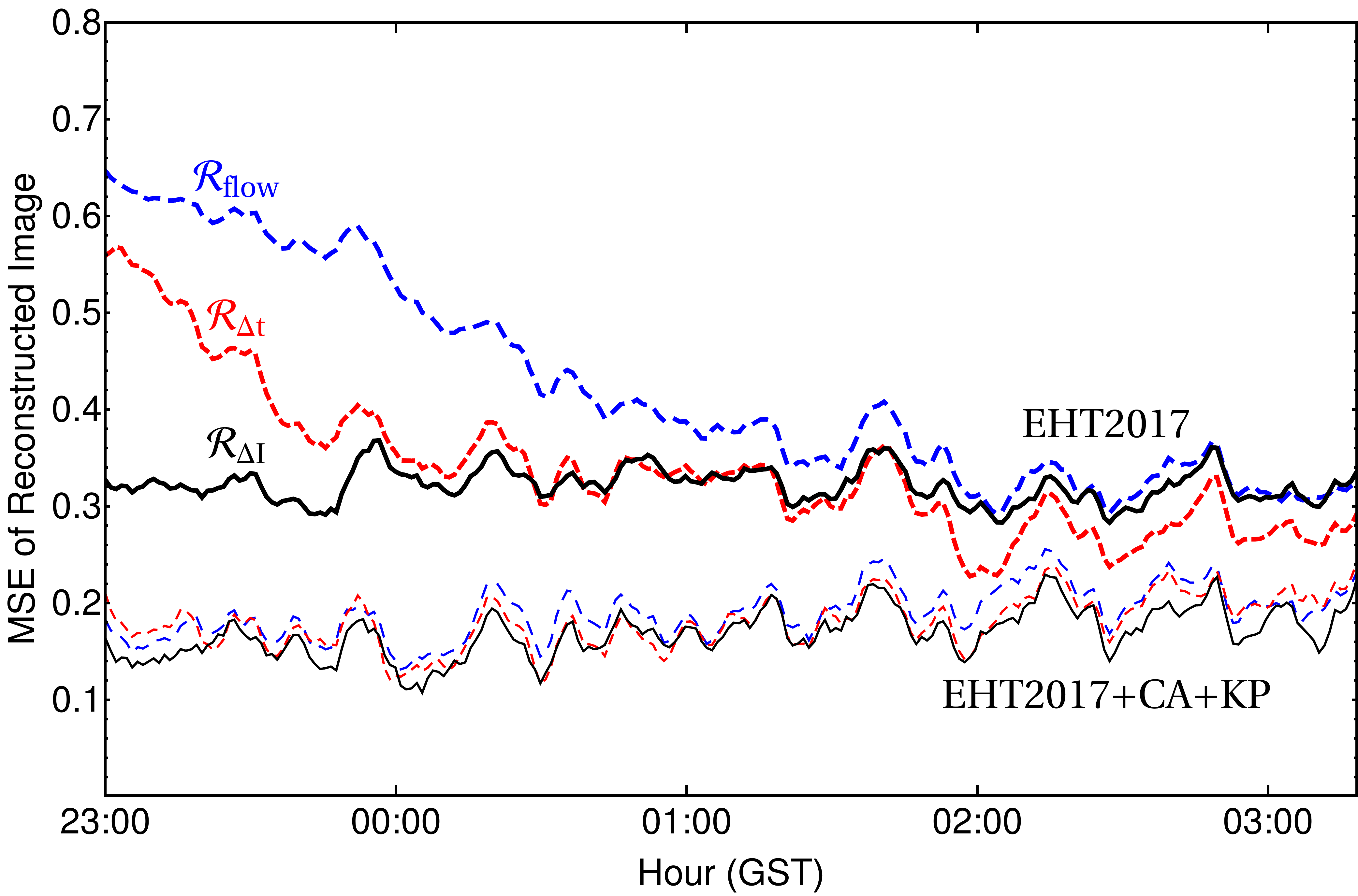}
\caption
{ 
MSE vs.\ time for the dynamical imaging reconstructions shown in Figures~\ref{fig::Method_Compare} and \ref{fig::Method_Compare_2}. Upper, thick lines show results for the 2017 EHT configuration; lower, thin lines include sites at the location of the CARMA array and at Kitt Peak. At early times, when the baseline coverage is minimal, $\mathcal{R}_{\Delta I}$ regularization provides the best results, showing the improvement that can be obtained under the assumption of a stable average image. At later times, $\mathcal{R}_{\Delta t}$ is as good, or slightly advantageous, showing the benefit when enforcing temporal continuity on reconstructed images. With the expanded EHT array configuration, all methods produce accurate and comparable results.
}
\label{fig::Method_Compare_MSE}
\end{figure}

\subsection{Comparison of Dynamical Imaging Methods}

We next compare all three dynamical imaging strategies on simulated observations of an accretion flow viewed at an inclination of $30^\circ$ with respect to the black hole's rotation axis. Apart from the viewing inclination, the simulation is identical to the one shown in Figure~\ref{fig::RdI_Example}. We used 1400 simulated movie frames, corresponding to 4.3 hours of observations, starting at a GST of 23:00. To simplify the comparison between the three imaging methods and avoid discrepancies from poor convergence or image misalignment, we used full complex visibilities for dynamical imaging and the $\mathcal{D}_2$ distance metric for each. Each reconstructed movie has 234 frames (1/6 the time resolution of the input movie).

Figure~\ref{fig::Method_Compare} compares the average image of the simulation with the averaged images from the three dynamical imaging reconstructions using the 2017 EHT configuration (for the reconstruction using $\mathcal{R}_{\rm flow}$ regularization, the reconstructed motion field is also shown). 
Figure~\ref{fig::Method_Compare_2} performs the same comparison for reconstructions that also included sites at Kitt Peak and at the location of the CARMA array.   Figure~\ref{fig::Method_Compare_MSE} shows the MSE over time for each reconstruction from both array configurations. Despite their differing assumptions about the underlying image variability, the three methods give results that are broadly consistent. With the additional EHT sites of the second example, the flow clearly identifies the correct direction of motion, although the estimated magnitude of the motion underestimates by a factor of several the estimated time-averaged optical flow of the simulated images \citep{Liu_2009}. Thus, to recover precise details about the motion will likely require either more stringent dynamical imaging constraints \citep[see, e.g.,][]{Bouman_2017_StarWarps} or additional sites added to the EHT.

\begin{figure*}[t]
\centering
\includegraphics*[width=0.245\textwidth]{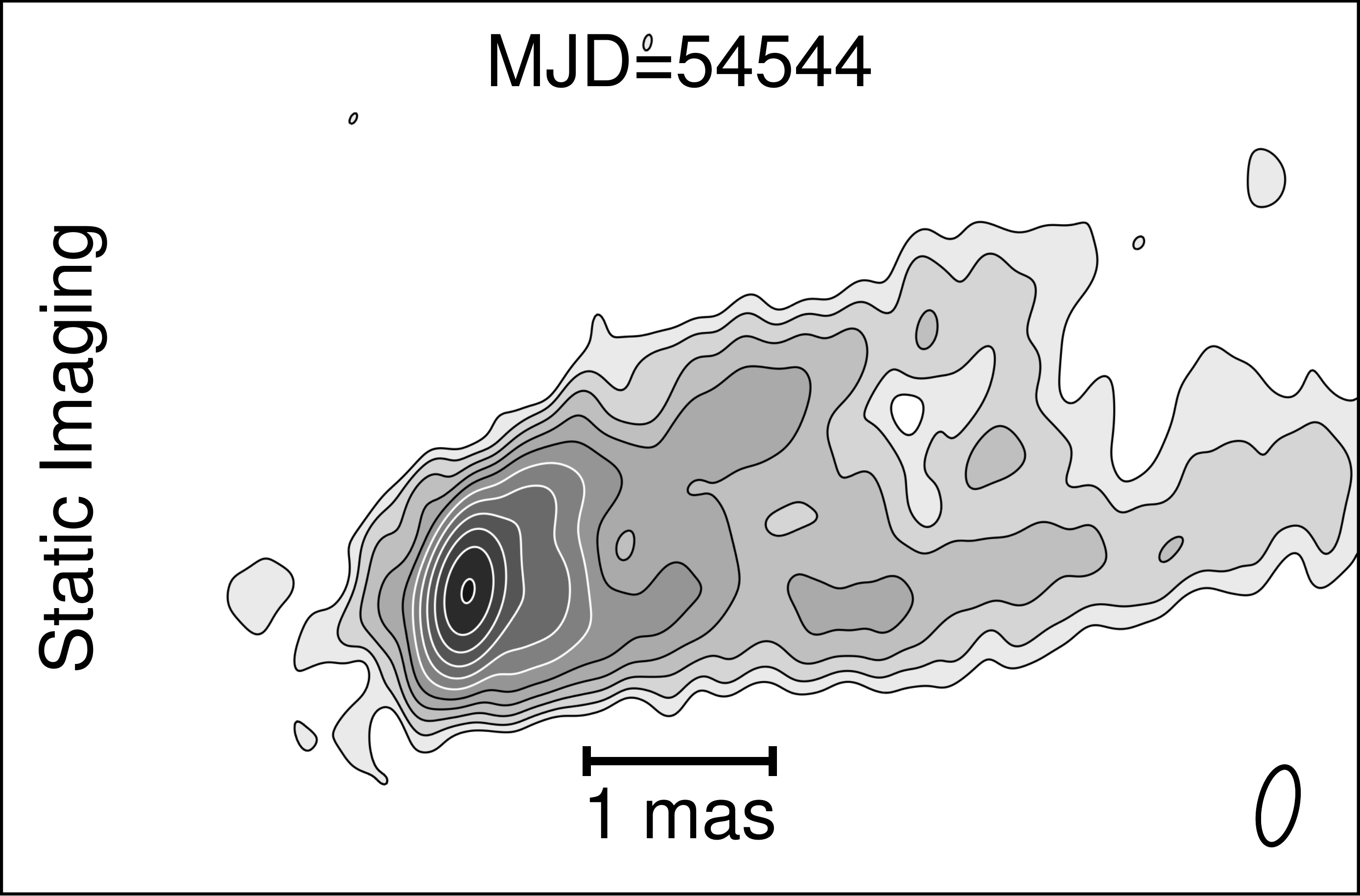}
\includegraphics*[width=0.245\textwidth]{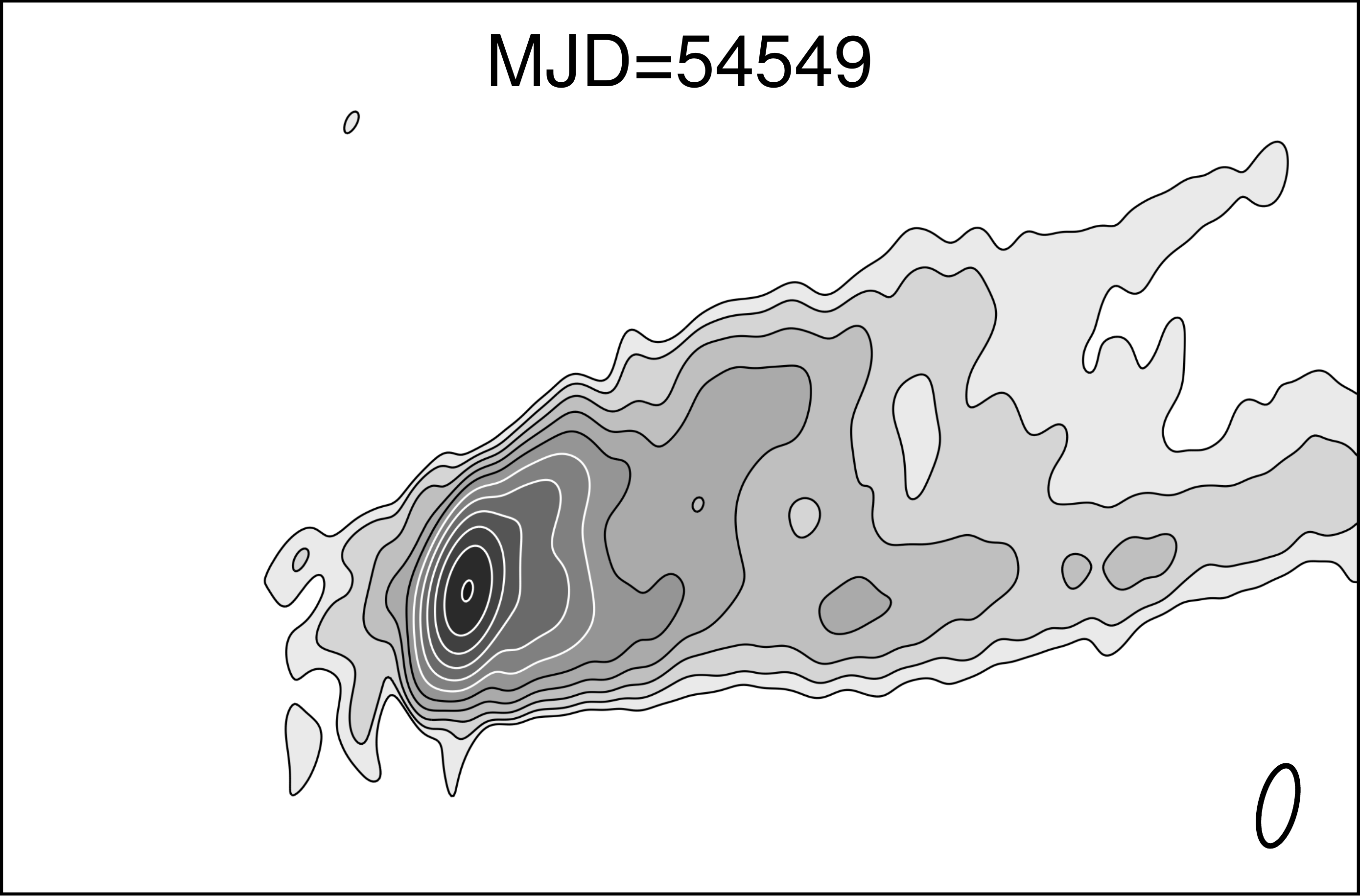}
\includegraphics*[width=0.245\textwidth]{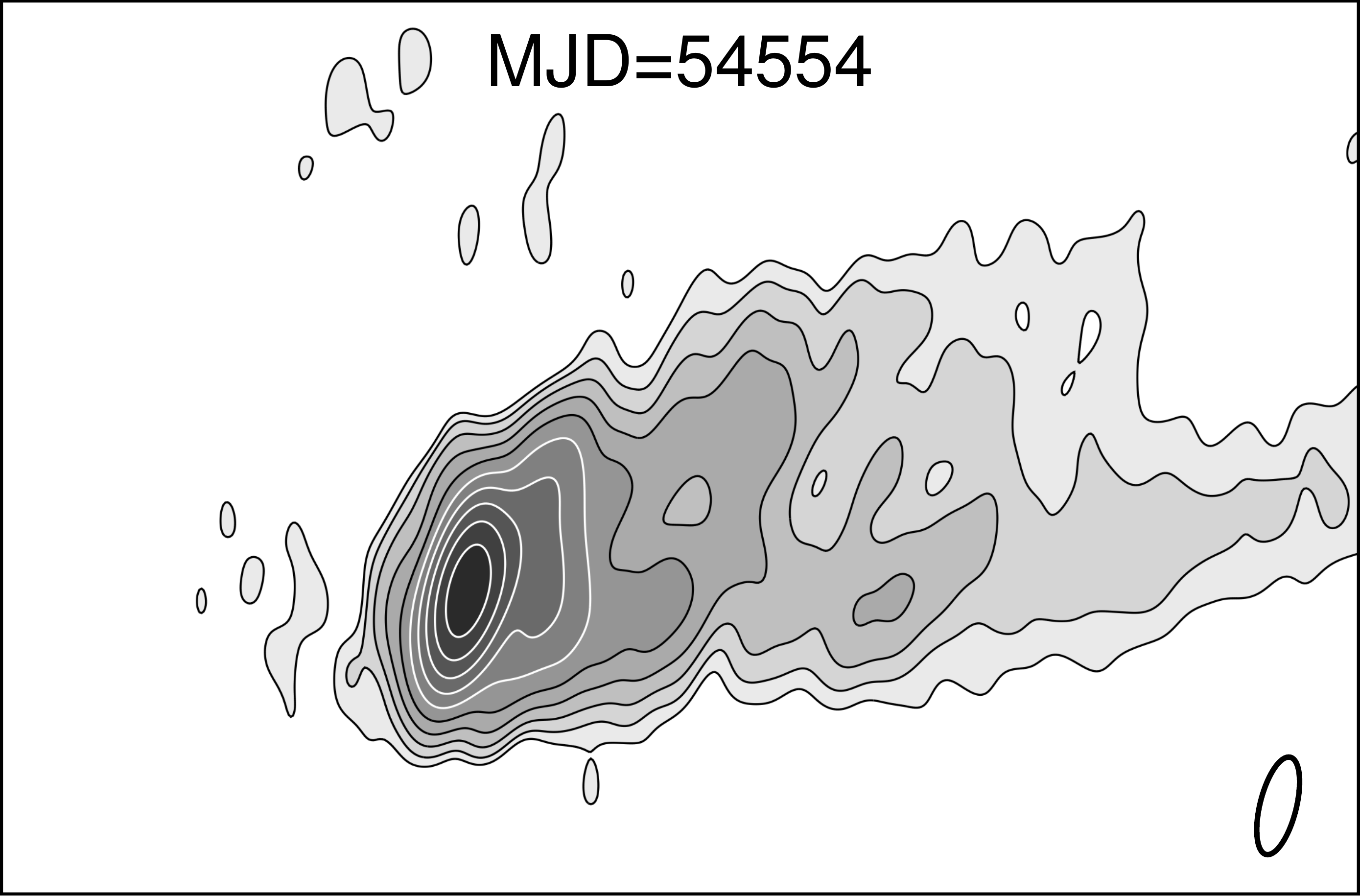}
\includegraphics*[width=0.245\textwidth]{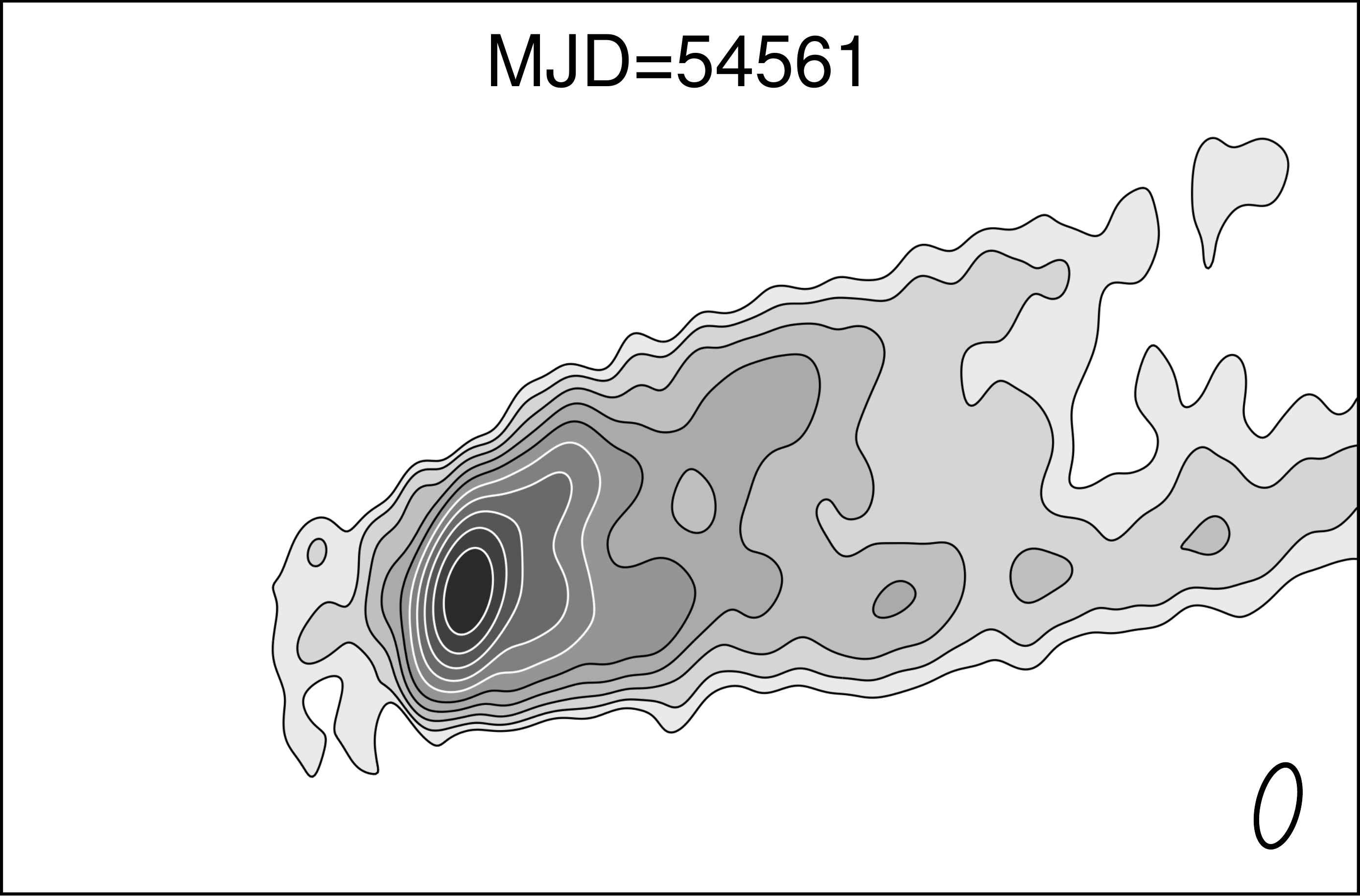}\\
\includegraphics*[width=0.245\textwidth]{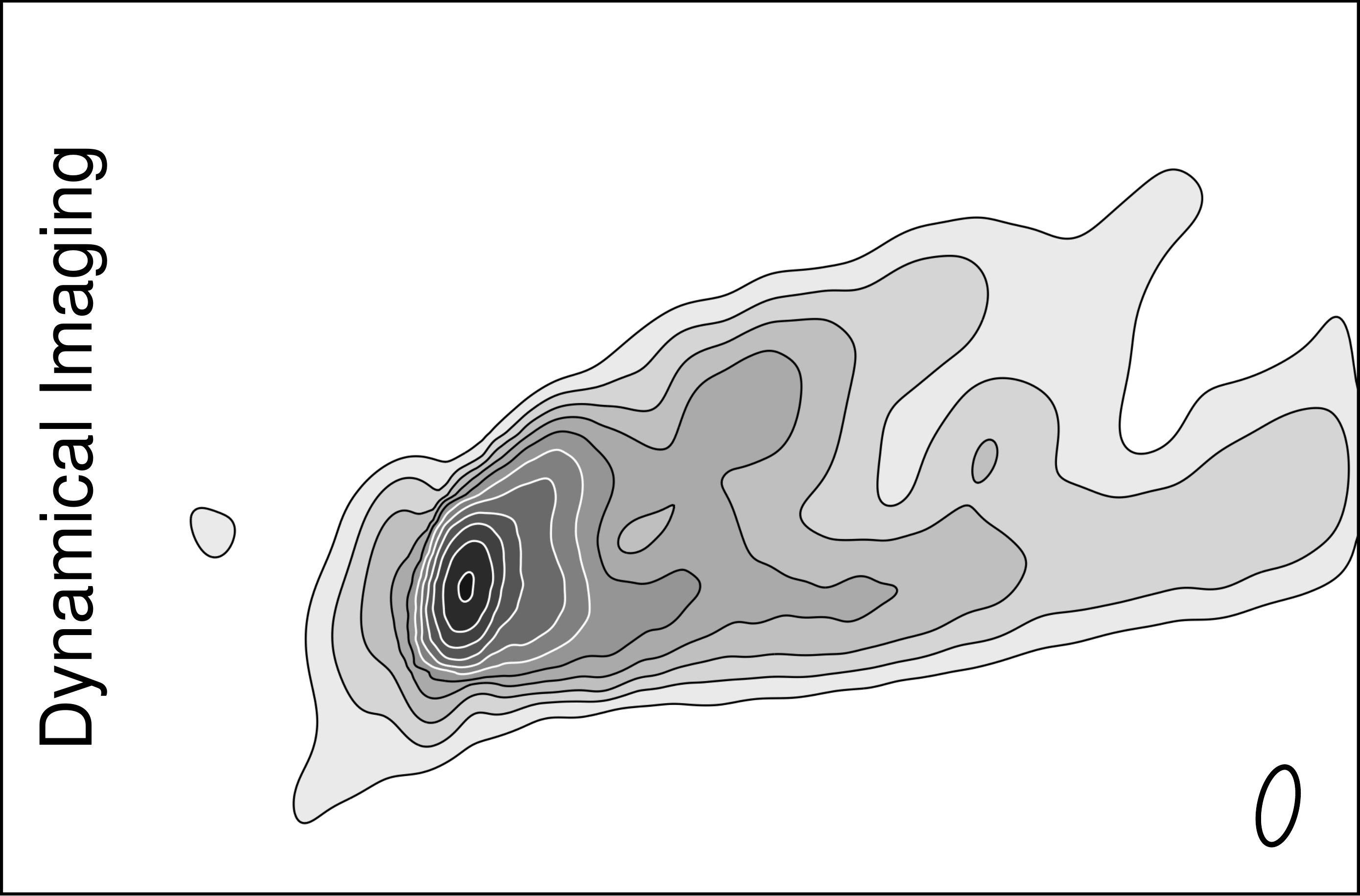}
\includegraphics*[width=0.245\textwidth]{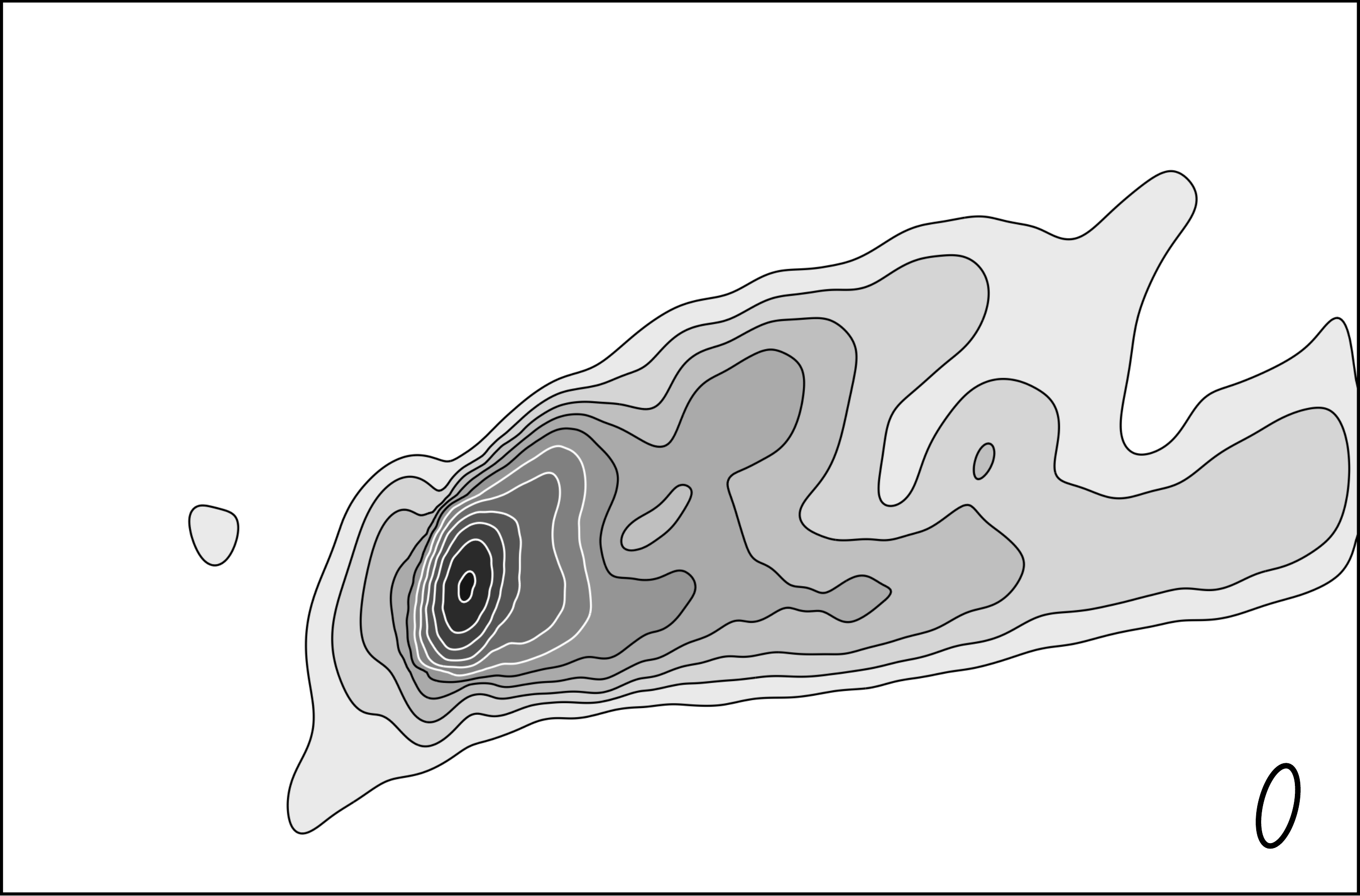}
\includegraphics*[width=0.245\textwidth]{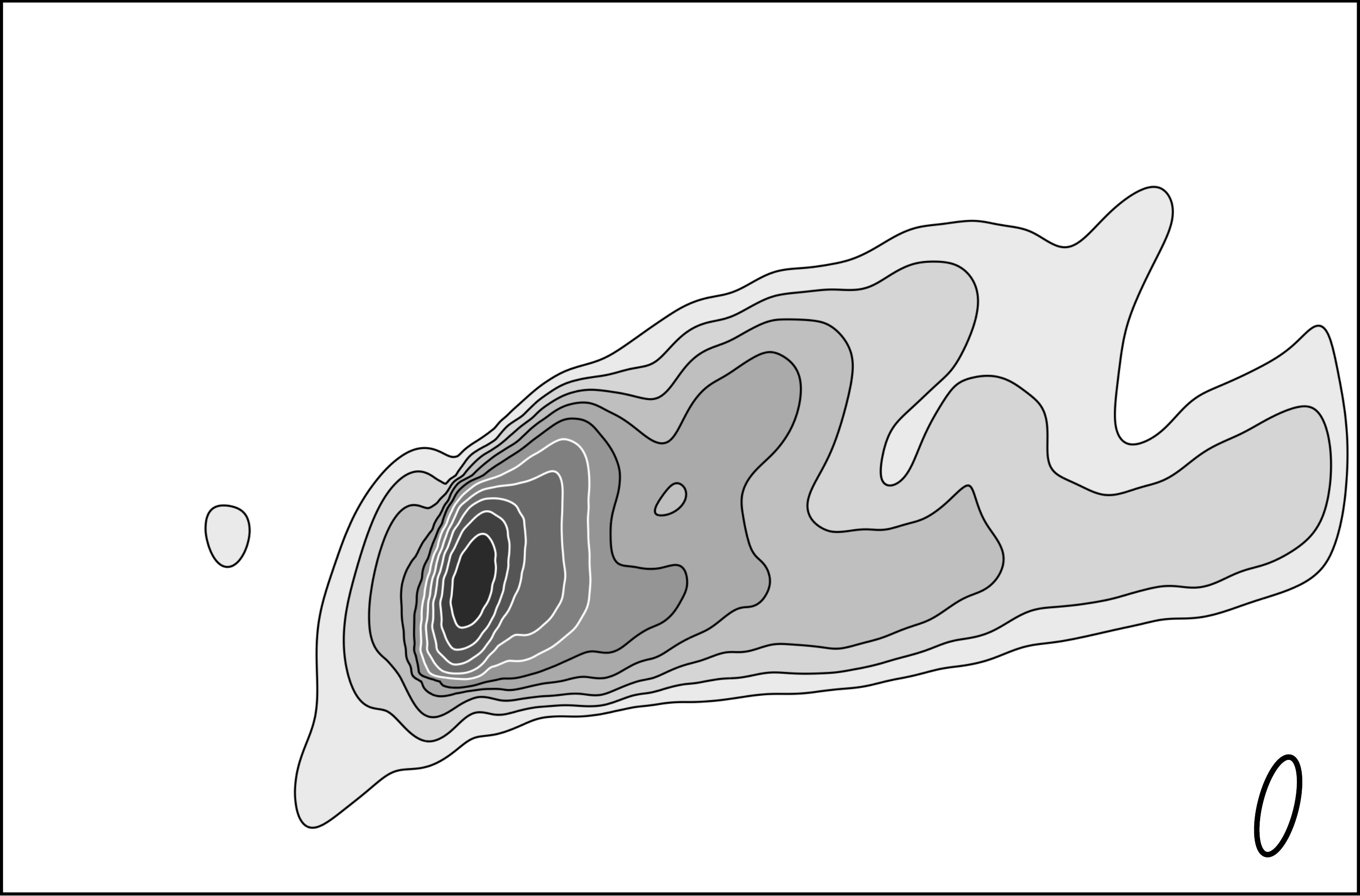}
\includegraphics*[width=0.245\textwidth]{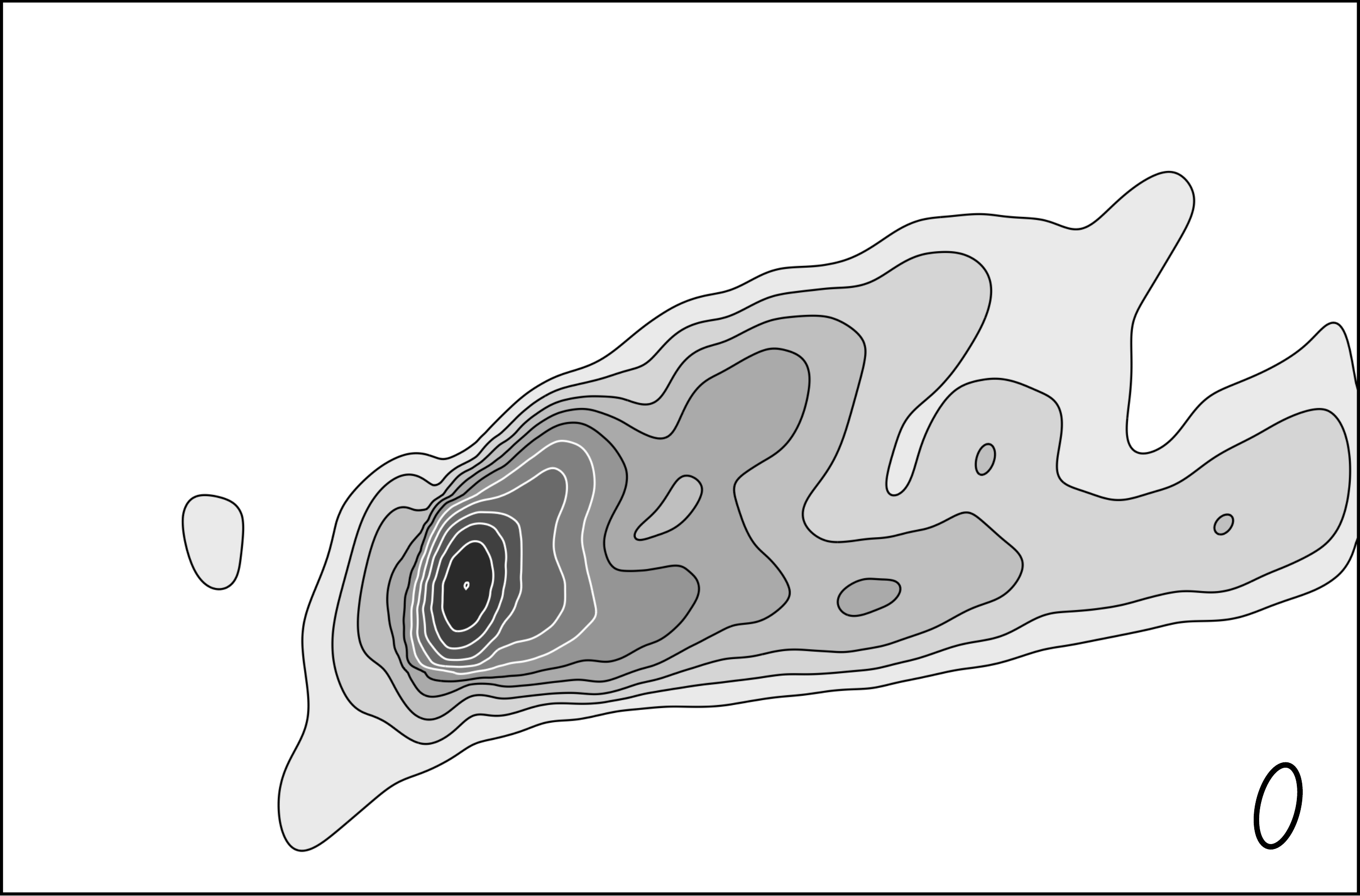}\\
\includegraphics*[width=0.245\textwidth]{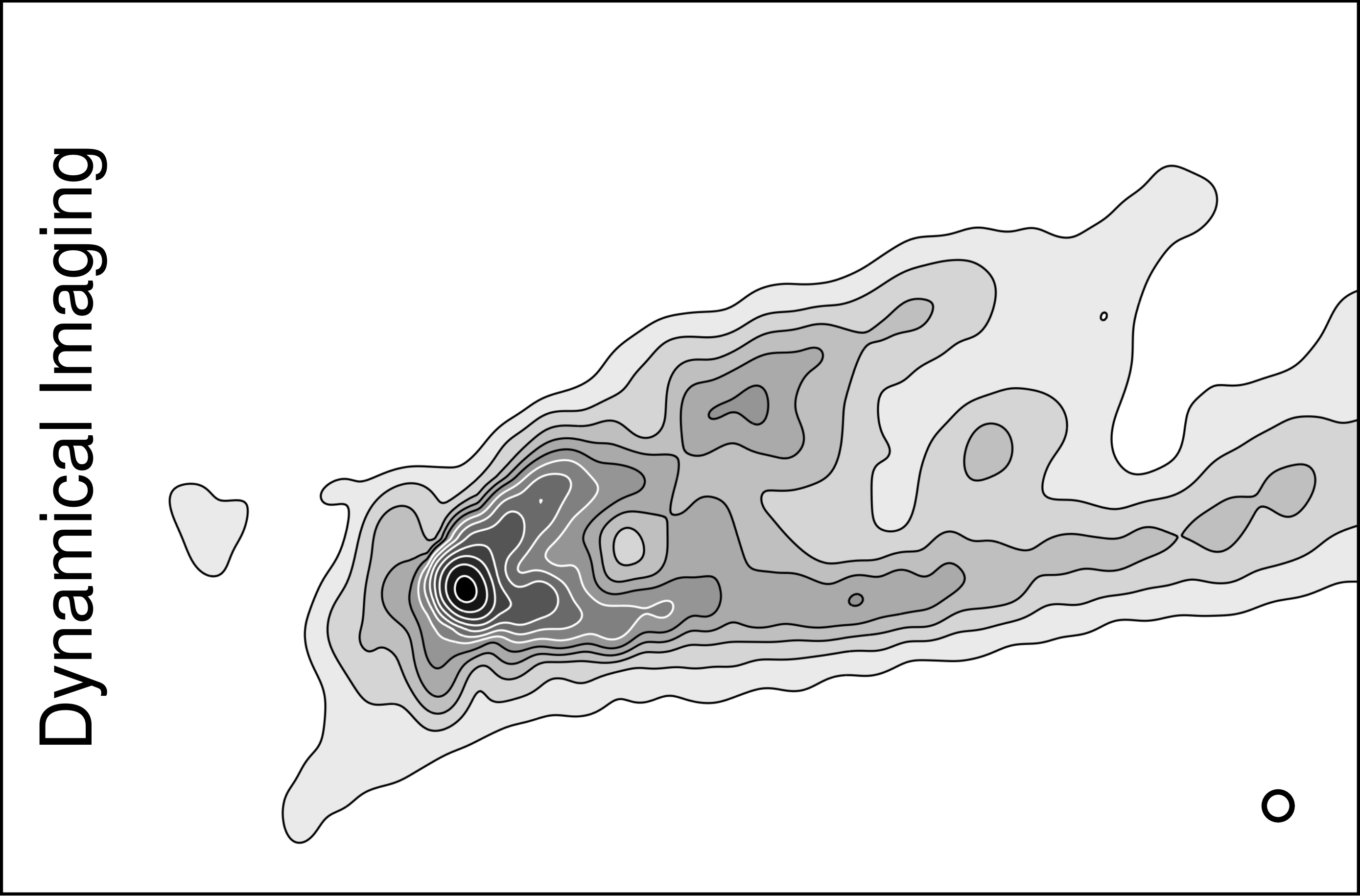}
\includegraphics*[width=0.245\textwidth]{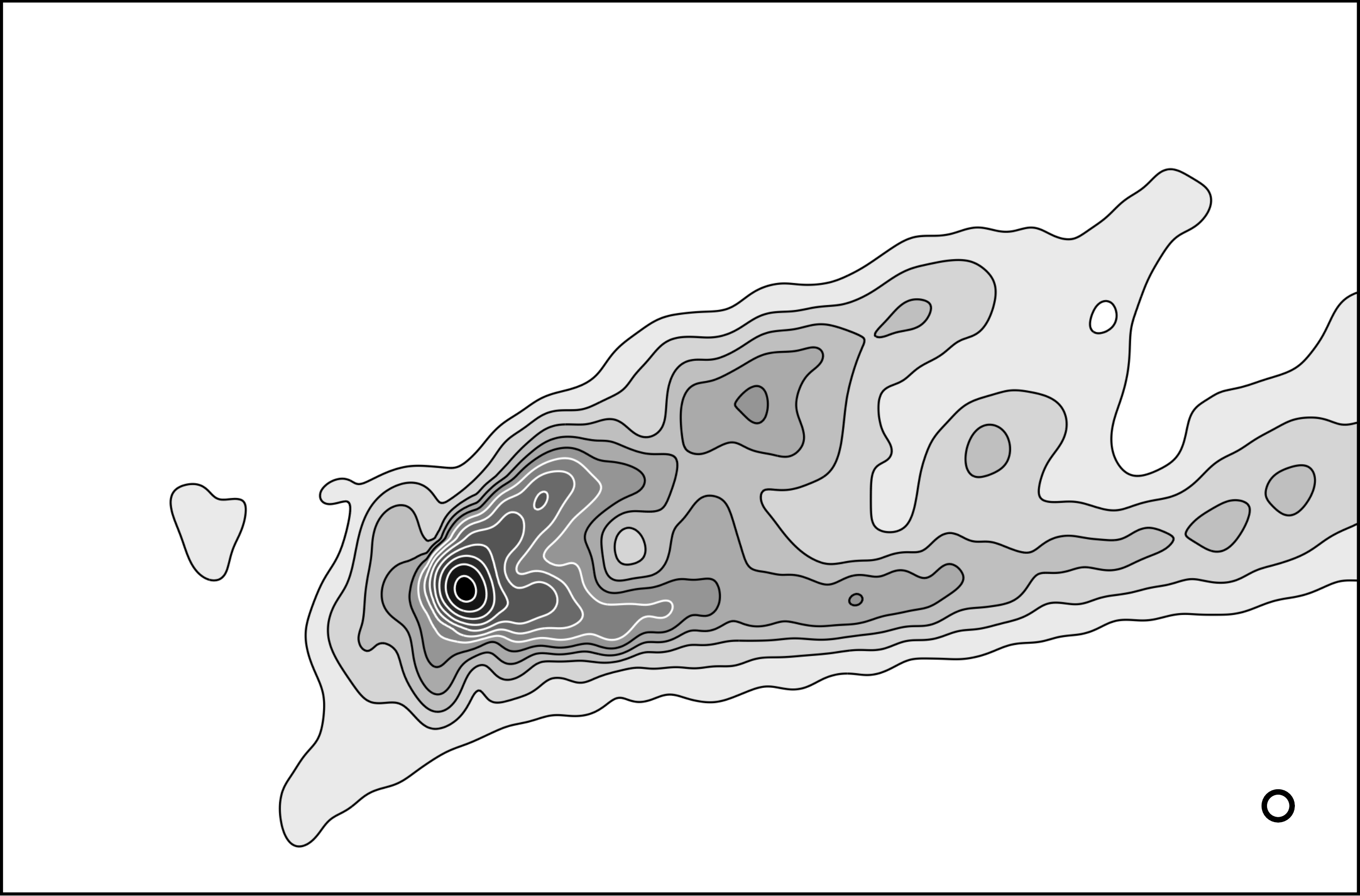}
\includegraphics*[width=0.245\textwidth]{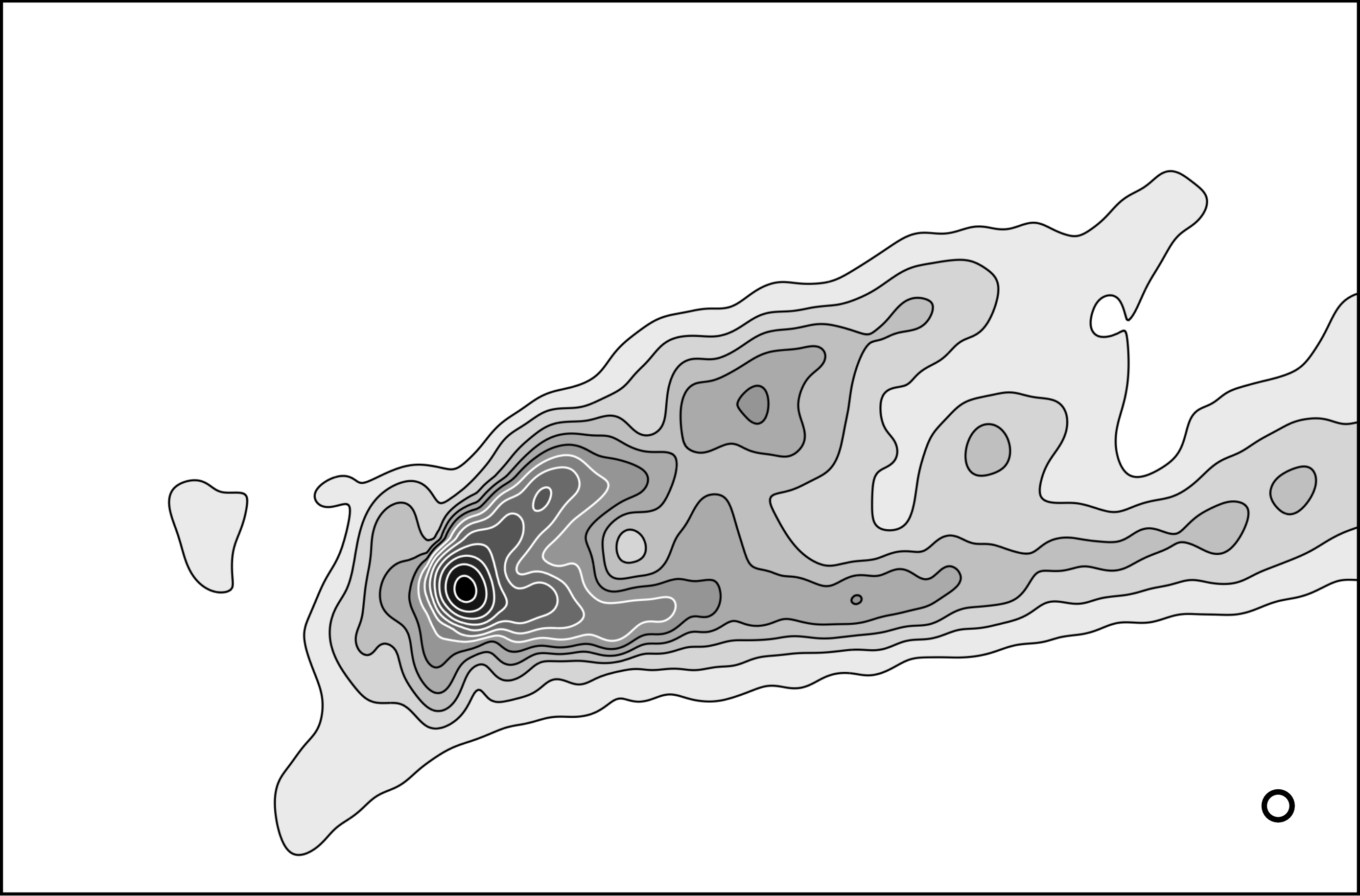}
\includegraphics*[width=0.245\textwidth]{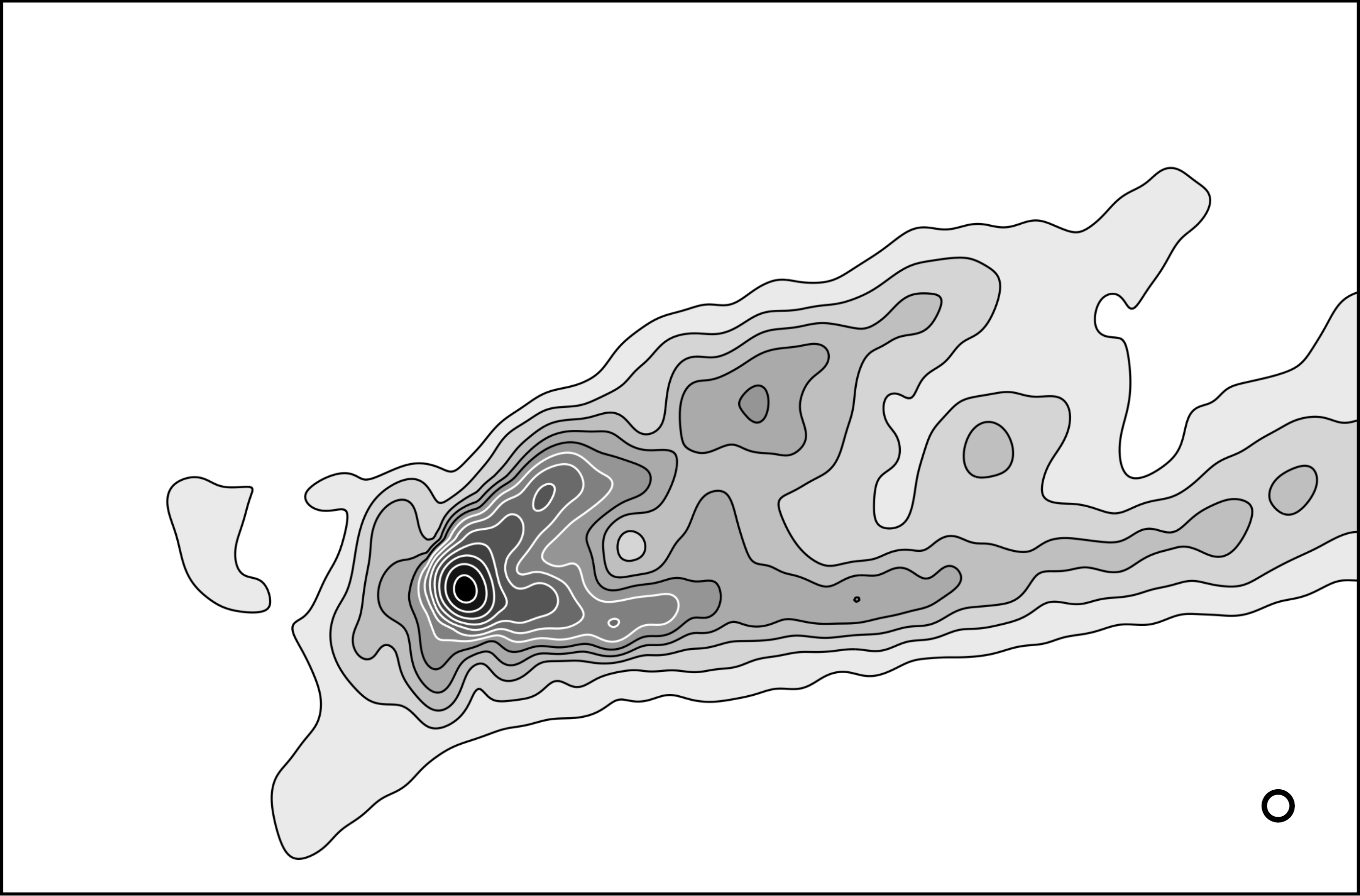}
\caption
{ 
Comparison of static imaging (top) and dynamical imaging (middle, bottom) of M87 for four closely spaced epochs over a span of 17 days. The static images were reconstructed using CLEAN \citep[for details, see][]{Walker_2016,Walker_2017}; the dynamical images were reconstructed using $\mathcal{R}_{\Delta t}$ regularization with symmetrized Kullback-Leibler divergence as the distance metric. The restoring beam of the CLEAN images varies from epoch to epoch but is typically ${\sim}430 \times 200~\mu{\rm as}$ at a position angle of $-13^\circ$. To simplify comparisons, the dynamical images in the middle row have been convolved with the corresponding CLEAN beam. To highlight evolution of compact structure, the dynamical images in the bottom row have been convolved with a circular Gaussian beam with FWHM of $150\,\mu{\rm as}$. At the resolution of the CLEAN beam, static and dynamical imaging are broadly consistent, but faint features are more similar from frame-to-frame in the dynamical reconstruction, more readily identifying physical evolution. Contours in all panels are at equal levels, starting at $10.5~{\rm mJy}/{\rm mas}^2$ (${=}1~{\rm mJy}/{\rm beam}$ in the first CLEAN image) and increasing by factors of 2.  
}
\label{fig::M87}
\end{figure*}

\subsection{Dynamical Imaging of M87}
\label{sec::M87}

As a final example, we used dynamical imaging on a series of 14 separate 43\,GHz VLBA observations of M87 taken over a span of 70 days in 2008 as part of the M87 Movie Project \citep{Walker_2016,Walker_2017}. We reconstructed a series of 24 images spaced by 3 days; each observation was associated to the nearest image in time, and the remaining images had no data constraints (see \S\ref{sec::Interpolation}). We used $\mathcal{R}_{\Delta t}$ regularization with symmetrized Kullback-Leibler divergence as the distance metric and $\sigma_{\Delta t} = 0$ (see \S\ref{sec::DiffGaussBlur}). We have found that this overall strategy works well for images with high dynamic range and irregular motion, without requiring fine tuning of the imaging parameters. We also utilized iterative self calibration, so that the dynamical imaging serves as both an imaging and calibration framework.

While detailed analysis of these results will be presented separately, Figure~\ref{fig::M87} shows four of the reconstructed frames with their corresponding static reconstructions over a short time interval (17 days). Even a moderately relativistic component, with an apparent transverse velocity of $2c$, would move by only $0.36~{\rm mas}$ over this entire interval.  Dynamical imaging successfully finds a series of similar images, each of which is consistent with its respective, self-calibrated data. By eliminating faint spurious features, outward motion along the jet is more readily evident in the dynamical reconstructions. Moreover, this interval includes one epoch (${\rm MJD}=54554$) for which the original data were adversely affected by poor weather and were not considered to be of adequate quality for inclusion in the final CLEAN dataset. Nevertheless, dynamical imaging is able to successfully link this period of inferior data to the higher quality nearby epochs so that the reconstructed image is not perceptibly degraded. 


\section{Summary}
\label{sec::Summary}

In summary, we have developed three regularizers that are suitable for dynamical imaging: $\mathcal{R}_{\Delta t}$, $\mathcal{R}_{\Delta I}$, and $\mathcal{R}_{\rm flow}$. $\mathcal{R}_{\Delta t}$ favors continuity from frame-to-frame, $\mathcal{R}_{\Delta I}$ favors frames that are small perturbations from the time-averaged image, and $\mathcal{R}_{\rm flow}$ favors frames that approximately evolve according to a stationary flow. For each of these regularizers, we have derived analytic gradients with respect to the unknown image parameters, so each converges quickly with modest computational resources (e.g., all the reconstructions in this paper were performed on a personal computer). Each can be used with any choice of VLBI data product (e.g., complex visibilities, the bispectrum, or visibility amplitudes and closure phases) and any choice or combination of image regularization for individual frames (e.g., entropy, $\ell_{n}$-norm, or total variation). 

For dynamical imaging, the most significant challenge we have encountered is suitable convergence to the optimal reconstruction, especially when using a small array and only VLBI closure quantities. We have discussed a number of strategies to assist convergence, most involving iterative re-imaging with blurring, averaging, individual frame imaging,  or other modifications at each stage. In \citet{Bouman_2017_StarWarps}, we develop an alternative approach to dynamical imaging that provides an analytic expression for the reconstructed images, lessening the problem of convergence, at the expense of restrictive constraints on the dynamical imaging framework. These methods can potentially be used in sequence to allow a flexible dynamical imaging strategy with reliable convergence. 

A major motivation for this work is the possibility of imaging \sgra\ with the EHT. Even in its simplest implementation, dynamical imaging provides a framework to estimate the time-averaged image of \sgra\ and will be significantly more sensitive than static imaging approaches if there is significant intrinsic variability (see Figure~\ref{fig::TimeAveragedImage}).   Dynamical imaging can also confirm key image features such as the black hole shadow based on their temporal signatures. For example, the region surrounding the shadow is expected to exhibit enhanced high-frequency variability \citep{Shiokawa_2017}. Dynamical studies of \sgra\ will be crucial for estimating accretion disk inclinations, breaking a degeneracy in time-averaged images from the near symmetry orthogonal to the rotation axis \citep[see, e.g.,][]{Broderick_2011,
Johnson_2015}, and they will also be helpful for estimating the black hole spin. Namely, while the shape of the black hole shadow is almost independent of spin \citep{Bardeen_1972,Takahashi_2004,Johannsen_2010}, orbital periods at the innermost stable circular orbit vary by nearly a factor of 10 depending on spin \citep{Bardeen_1973}.  

For EHT imaging, $\mathcal{R}_{\Delta I}$ regularization appears especially promising because inhomogeneous data can be combined regardless of their spacing, specific observing cadence, or participating telescopes. Thus, any robust data products from multiple epochs can be merged to produce an average image that is not adversely affected by the variability, while also estimating the time-dependent perturbations. Moreover, our approach can be combined with other regularization frameworks, for instance to simultaneously mitigate the effects of interstellar scattering \citep{Stochastic_Optics}. Looking forward, dynamical imaging may be a rich source of continued study of \sgra\ as millimeter VLBI continues to expand, potentially even to Earth-space baselines \citep[see, e.g.,][]{Wild_2009,Smirnov_2012,Kardashev_2014}.

Our framework is also suitable for multi-epoch VLBI imaging studies, including kinematical studies of relativistic jets \citep[e.g.,][]{Kellermann_2004,Jorstad_2005,Lister_2009,Walker_2016,Mertens_Lobanov_2016,Lister_2016}, supernovae \citep[e.g.,][]{Bartel_2000,Bietenholz_2003,Bartel_2009}, and microquasars \citep[e.g.,][]{Fomalont_2001,Mioduszewski_2004,Jeffrey_2016}. It can also be applied to multi-epoch wide-field imaging, where highly sensitive instruments such as the Square Kilometer Array (SKA) are expected to detect a combination of static and variable sources \citep[e.g.,][]{Metzger_2015,Fender_2015}.  For cases with regular motion, dynamical imaging with $\mathcal{R}_{\rm flow}$ can self-consistently estimate the velocity field, while cases with irregular motion should be imaged with softer regularization, such as $\mathcal{R}_{\Delta I}$, $\mathcal{R}_{\Delta t}$ (see Figure~\ref{fig::M87}), or a non-stationary flow.  A benefit of our approach is that epochs with poor u-v coverage or sensitivity can partially inherit the higher resolution of neighboring epochs -- our approach does not assume a constant VLBI beam among the epochs or even a constant spacing between epochs (see, e.g., Figure~\ref{fig::RdI_Example_MSE}). Our approach can also incorporate iterative self-calibration to derive a calibration solution that is compatible with smooth structural evolution among epochs. And while our focus has been on dynamical imaging of the total flux density, our methods are straightforward to adapt to full-Stokes polarization, which often shows more pronounced variability than the total flux density \citep[e.g.,][]{Gabuzda_2000}.

\acknowledgements{We are particularly indebted to Craig Walker for sharing his calibrated M87 data and images and for many illuminating discussions. We gratefully acknowledge helpful conversations with John Wardle, James Guillochon, and Maciek Wielgus. We also thank Alan Marscher, Svetlana Jorstad, Dan Homan, and Matt Lister for making their multi-epoch VLBI studies available to help refine and test our code. We thank Avery Broderick for providing the hot spot simulations used in this work. We thank the referee for suggesting the application of this technique to wide-field imaging studies with next generation arrays. 
We thank the National Science Foundation (AST-1440254, AST-1614868) and the Gordon and Betty Moore Foundation (GBMF-3561, GBMF-5278) for financial support of this work. This work was supported in part by the Black Hole Initiative at Harvard University, which is supported by a grant from the John Templeton Foundation. F.R.\ is supported by the ERC Synergy Grant ``BlackHoleCam'' (Grant 610058). K.A.\ is financially supported by the program of Postdoctoral Fellowships for Research Abroad at the Japan Society for the Promotion of Science (JSPS).}

\begin{appendix}

\section{Gradients of Dynamical Regularization Terms}

We will now derive analytic expressions for the gradients of the dynamical regularization terms. These gradients depend on the chosen distance function, and so we will provide representative examples. 
 
\subsection{Gradients of $\mathcal{R}_{\Delta t}$}

The gradient of $\mathcal{R}_{\Delta t}$ when using the $\mathcal{D}_{2}$ distance function is given by
\begin{align}
\label{eq::Rdt_Gradient}
\frac{\partial{\mathcal{R}_{\Delta t}}}{\partial \mathbf{I}_{k}} &= 2 B^2\left( \left[\mathbf{I}_{k} - \mathbf{I}_{k-1} \right] \delta_{k>1} + \left[\mathbf{I}_k - \mathbf{I}_{k+1} \right]\delta_{k<N_{\rm t}} \right),
\end{align}
where $B^2$ denotes a blurring operator that is applied twice, and the indicator function $\delta_x$ is defined to be unity when the subscripted condition $x$ is satisfied and is zero otherwise. Note that $B$ could be replaced by any $N^2 \times N^2$ matrix operator $\lrover{\mathbf{B}}$ (acting on an image vector), in which case $B^2$ in Eq.~\ref{eq::Rdt_Gradient} must be replaced by $\lrover{\mathbf{B}}^\intercal \lrover{\mathbf{B}}$. 

Likewise, for the $\mathcal{D}_{p}$ distance function, 
\begin{align}
\label{eq::Rdt_Gradient_wp}
\frac{\partial{\mathcal{R}_{\Delta t}}}{\partial \mathbf{I}_{k}} &= p B\left( \left|B\left(\mathbf{I}_{k} - \mathbf{I}_{k-1} \right)\right|^{p-1} \mathrm{sgn}\left(B\left(\mathbf{I}_{k} - \mathbf{I}_{k-1} \right) \right) \delta_{k>1} + \left|B\left(\mathbf{I}_{k} - \mathbf{I}_{k+1} \right)\right|^{p-1} \mathrm{sgn}\left(B\left(\mathbf{I}_{k} - \mathbf{I}_{k+1} \right) \right) \delta_{k<N_{\rm t}}      \right).
\end{align}

Lastly, for the $\mathcal{D}_{\rm KL}$ distance function, 
\begin{align}
\label{eq::Rdt_Gradient_KL_Appendix}
\frac{\partial{\mathcal{R}_{\Delta t}}}{\partial \mathbf{I}_{k}} &= B\left( \left[ \mathbf{1} + \ln \frac{ B(\mathbf{I}_{k})}{B(\mathbf{I}_{k-1})}\right]\delta_{k>1} - \frac{B(\mathbf{I}_{k+1})}{B(\mathbf{I}_{k})} \delta_{k<N_{\rm t}} \right),
\end{align}
where $\mathbf{1}$ denotes a vector of length $N^2$ with every element equal to unity. Gradients for variants of the $\mathcal{D}_{\rm KL}$ function can be computed similarly. We again emphasize that operations such as norms ($|\dots|$), quotients, powers, and products of image vectors are to be computed elementwise.  

\subsection{Gradients of $\mathcal{R}_{\Delta I}$}
For the $\mathcal{R}_{\Delta I}$ regularization function, using the $\mathcal{D}_2$ distance function gives the following gradient:
\begin{align}
\frac{\partial \mathcal{R}_{\Delta I}}{\partial \mathbf{I}_k} &= 2 \left( \mathbf{I}_k - \mathbf{I}_{\rm avg} \right).
\end{align}

For the $\mathcal{D}_p$ distance function, we obtain
\begin{align}
\frac{\partial \mathcal{R}_{\Delta I}}{\partial \mathbf{I}_k} &= p \left| \mathbf{I}_{k} - \mathbf{I}_{\rm avg} \right|^{p-1} \mathrm{sgn}\left( \mathbf{I}_{k} - \mathbf{I}_{\rm avg} \right) - \frac{p}{N_{\rm t}} \sum_{j=1}^{N_{\rm t}}  \left| \mathbf{I}_{j} - \mathbf{I}_{\rm avg} \right|^{p-1} \mathrm{sgn}\left( \mathbf{I}_{j} - \mathbf{I}_{\rm avg} \right).
\end{align}
Note that the second term is independent of $k$ and is zero for $p=2$. 

Lastly, for the $\mathcal{D}_{\rm KL}$ distance function, we find
\begin{align}
\frac{\partial \mathcal{R}_{\Delta I}}{\partial \mathbf{I}_k} &= 1 - \frac{\mathbf{I}_{\rm avg}}{\mathbf{I}_{k}} + \frac{1}{N_{\rm t}} \sum_{j=1}^{N_{\rm t}} \ln\left( \frac{\mathbf{I}_{\rm avg}}{\mathbf{I}_{j}} \right).
\end{align}

\subsection{Gradients of $\mathcal{R}_{\rm flow}$}

For $\mathcal{R}_{\rm flow}$ regularization, we must evaluate the gradients of $\mathcal{R}_{\rm flow}$ with respect to both the images and the flow. The gradient with respect to the images can be written in a general form that only depends on the linear operator $\lrover{\mathbf{F}}_{\rm flow}$ and its transpose. For instance, using the $\mathcal{D}_2$ distance metric gives 
\begin{align}
\label{eq::dRflow_dI_prep}
\frac{\partial \mathcal{R}_{\rm flow}}{\partial \mathbf{I}_k} &= 2 \left( \mathbf{I}_{k} - \lrover{\mathbf{F}}_{\rm flow} \cdot \mathbf{I}_{k-1}\right) \delta_{k>1} - 2\lrover{\mathbf{F}}_{\rm flow}^\intercal \cdot \left( \mathbf{I}_{k+1} - \lrover{\mathbf{F}}_{\rm flow} \cdot \mathbf{I}_k \right) \delta_{k<N_{\rm t}}.
\end{align}
Using the identify that $\nabla^\intercal = - \nabla$ (appropriate for central finite difference operators), we obtain $(\mathbf{m} \cdot \nabla)^\intercal = -\left( \nabla \cdot \mathbf{m}  + \mathbf{m} \cdot \nabla \right)$. The other elements of $\lrover{\mathbf{F}}_{\rm flow}$ are diagonal matrices, so
\begin{align}
\lrover{\mathbf{F}}_{\rm flow}^\intercal &= \left( 1 - \mathbf{m} \cdot \nabla - \nabla \cdot \mathbf{m} \right)^\intercal\\
\nonumber &= 1 + \left( \nabla \cdot \mathbf{m}  + \mathbf{m} \cdot \nabla \right) - \nabla \cdot \mathbf{m} \\
\nonumber &= 1 + \mathbf{m} \cdot \nabla. 
\end{align}
Substituting this result into Eq.~\ref{eq::dRflow_dI_prep}, we obtain
\begin{align}
\label{eq::dRflow_dI}
\frac{\partial \mathcal{R}_{\rm flow}}{\partial \mathbf{I}_k} &= 2 \left( \mathbf{I}_{k} - \lrover{\mathbf{F}}_{\rm flow} \cdot \mathbf{I}_{k-1}\right) \delta_{k>1} - 2 \left(  1 + \mathbf{m}\cdot \nabla \right) \cdot \left( \mathbf{I}_{k+1} - \lrover{\mathbf{F}}_{\rm flow} \cdot \mathbf{I}_k \right) \delta_{k<N_{\rm t}}.
\end{align}
Likewise, the gradient with respect to the flow vector field $\mathbf{m}$ is given by
\begin{align}
\frac{\partial \mathcal{R}_{\rm flow}}{\partial \mathbf{m}} &= \sum_{j=1}^{N_{\rm t}-1}\left[ 2 \left( \mathbf{I}_{j+1} - \lrover{\mathbf{F}}_{\rm flow} \mathbf{I}_{j}\right) \nabla\mathbf{I}_j  - \nabla \left( 2 \left( \mathbf{I}_{j+1} - \lrover{\mathbf{F}}_{\rm flow} \cdot \mathbf{I}_{j}\right) \mathbf{I}_{j} \right) \right]\\ 
\nonumber  &\approx -2\sum_{j=1}^{N_{\rm t}-1} \mathbf{I}_j \nabla \left( \mathbf{I}_{j+1} - \lrover{\mathbf{F}}_{\rm flow} \cdot \mathbf{I}_{j}\right).
\end{align}

\end{appendix}

\ \\

\bibliography{Dynamical_Imaging.bib}

\end{document}